\documentclass[12pt]{article}
\usepackage{amstext}
\usepackage{amssymb}
\usepackage{amsmath}
\usepackage{latexsym}

\catcode `\@=11 \@addtoreset{equation}{section}

\catcode `\@=12

\newcommand{\R}{R \!\!\!\! R}

\newcommand{\1}{1 \!\!\! 1}

\newtheorem{defi}{Definition}[section]
\newtheorem{lem}[defi]{Lemma}
\newtheorem{prop}[defi]{Proposition}
\newtheorem{theorem}[defi]{Theorem}
\newtheorem{cor}[defi]{Corollary}

\newcommand{\bedefin}{\begin{defi}}
\newcommand{\findefi}{\end{defi} \medskip}

\newcommand{\betheo}{\begin{theorem}$\!\!${\bf \,\,\,}}
\newcommand{\entheo}{\end{theorem}}
\newcommand{\enth}{\end{theorem}}

\newcommand{\becor}{\begin{cor}$\!\!${\bf .}}
\newcommand{\encor}{\end{cor}}

\newcommand{\belem}{\begin{lem}$\!\!${\bf .}}
\newcommand{\enlem}{\end{lem}}

\newcommand{\beprop}{\begin{prop}$\!\!${\bf .}}
\newcommand{\enprop}{\end{prop}}

\newcommand{\be}{\begin{equation}}
\newcommand{\en}{\end{equation}}
\newcommand{\ee}{\end{equation}}

\newcommand{\bea}{\begin{eqnarray}}
\newcommand{\ena}{\end{eqnarray}}

\newcommand{\beano}{\begin{eqnarray*}}
\newcommand{\enano}{\end{eqnarray*}}

\newcommand{\bee}{\begin{enumerate}}
\newcommand{\ene}{\end{enumerate}}

\newcommand{\bei}{\begin{itemize}}
\newcommand{\eni}{\end{itemize}}

\newcommand{\betab}{\begin{tabular}}
\newcommand{\entab}{\end{tabular}}

\newcommand{\bd}{\begin{displaymath}}

\newcommand{\h}{{\mathfrak H}}
\newcommand{\kk}{{\mathfrak K}}

\newcommand{\htil}{\widetilde{\mathfrak H}}

\newcommand{\bvarepsilon}{\mbox{\boldmath $\varepsilon$}}

\newcommand{\bgamma}{\mbox{\boldmath $\gamma$}}

\newcommand{\bPhi}{\mbox{\boldmath $\Phi$}}

\newcommand{\bK}{\mathbf K}

\newcommand{\bJ}{\mathbf J}

\def\H{\relax\ifmmode {\mathcal H}\else${\mathcal H}$\fi}

\newcommand{\Leb}[2]{
  \ifnum #2=1
  L^#1(\R,\,\ud x)
  \fi
  \ifnum #2=2
  L^#1(\R^2,\,\ud^2 \vec{x}\,)
  \fi}

\newcommand{\ud}{\mathrm{d}}

\begin{document}

\thispagestyle{empty}

\vspace*{1cm}

\begin{center}
{\Large \bf Some Physical Appearances of \\
Vector Coherent States and CS Related to \\[3mm]
 Degenerate Hamiltonians} \vspace{.75cm}\\

{\large S. Twareque Ali}
\footnote[1]{Department of Mathematics and Statistics, Concordia University,
Montr\'eal, Qu\'ebec, CANADA H4B 1R6\\
e-mail: stali@mathstat.concordia.ca}
\vspace{3mm}\\

{\large F. Bagarello}
\footnote[2]{ Dipartimento di Matematica ed Applicazioni,
Facolt\`a di Ingegneria, Universit\`a di Palermo, I-90128  Palermo, ITALY\\
e-mail: bagarell@unipa.it\,\,\,\, Home page: www.unipa.it/$^\sim$bagarell}
\end{center}

\vspace*{.50cm}

\begin{abstract}
\noindent In the spirit of some earlier work on the construction
of vector coherent states over matrix domains, we compute here
such states associated to some physical Hamiltonians. In
particular, we construct vector coherent states of the
Gazeau-Klauder type. As a related problem, we also  suggest a way
to handle degeneracies in the Hamiltonian for building coherent
states. Specific physical Hamiltonians studied include a single
photon mode interacting with a pair of fermions, a Hamiltonian
involving a single boson and a single fermion, a charged particle
in a three dimensional harmonic force field  and the case of a
two-dimensional electron placed in a constant magnetic field,
orthogonal to the plane which contains the electron. In this last
example, an interesting modular structure emerges for two underlying
von Neumann algebras, related to opposite directions of the
magnetic field. This leads to the existence of coherent states
built out of KMS states for the system.

\end{abstract}

\bigskip

 \vfill

\newpage

\section{Introduction}
   In some earlier work \cite{AEG,sanali}, a fairly
systematic method has been  introduced for constructing vector
coherent states over various types of matrix domains. The
construction included earlier types of vector coherent states,
arising  mainly in nuclear physical problems, under the additional
assumption of the existence of a resolution of the identity. (A
detailed discussion of this point, as well as an exhaustive
reference to the earlier literature is given in \cite{AEG}). In
the present paper we apply the method developed in
\cite{AEG,sanali} to construct vector coherent states arising from
various physical Hamiltonians. The kind of coherent states we
generate are thus vectorial generalizations of the Gazeau-Klauder
type \cite{GK} of coherent states. Some of the Hamiltonians we
consider have degenerate spectra and in order to deal with this
situation, we attempt a second generalization of the
Gazeau-Klauder formalism. There have been earlier attempts in the
literature for handling degeneracies when constructing coherent
states associated to Hamiltonians \cite{fox,klau}. The method we
suggest here is somewhat different from the one suggested in
\cite{klau} and radically different from that suggested in
\cite{fox}. However, we feel that the present method is more
economical in the introduction of additional parameters defining
the coherent states -- we only need one additional parameter. We
also look at situations where the degeneracy is countably
infinite. In this context, in the case of a two-dimensional
electron placed in a constant magnetic field, orthogonal to the
plane which contains the electron, we encounter a highly
interesting modular algebraic structure generated by the
observables of the problem, leading to the rather unexpected
appearance of equilibrium statistical mechanical states of the
well-known KMS type \cite{hahuwi}. It is worth recalling that this
model is quite an interesting one, since it is the building block
for writing down the many-body Hamiltonian of the fractional
quantum Hall effect, see \cite{BMS} and references therein. It is
well known that the eigenspectrum of the single electron
Hamiltonian can be found explicitly, and that there exists an infinite
degeneracy for each eigenvalue (the so-called {\em Landau levels})
\cite{bagant}.

   The rest of this paper is organized as follows: In Section \ref{GK-revis} we
review the Gazeau-Klauder construction within the framework of reproducing kernel Hilbert spaces.
This general framework is then used in Section \ref{GK-VCS} to construct vector coherent
states of the Gazeau-Klauder type. We illustrate the method with a couple of physical
examples. Section \ref{ham-degen} generalizes the treatment to Hamiltonians with degeneracies. We
treat the cases of finite and infinite degeneracies separately and illustrate the finite situation
with a number of physical examples. In Section \ref{elec-mag-fld} we work out, in detail,
a physical example in which infinite degeneracies occur. In this example we also
observe the existence of a modular algebraic structure and the appearance of KMS
states, familiar from equilibrium statistical mechanics. Finally, in the Appendix we collect
together explicit computations of some of the more unfamiliar formulae in Sections
\ref{fin-deg} and \ref{elec-mag-fld}.

\section{The Gazeau-Klauder scheme revisited}\label{GK-revis}

The Gazeau-Klauder scheme  \cite{GK} is a method for constructing  coherent states
$|J,\gamma\rangle$, where $J\geq 0$ and $\gamma\in\mathbb{R}$, associated to
physical Hamiltonians $H$, which have discrete non-degenerate spectra. The states have to
satisfy to the following properties:

\begin{itemize}
\item {\em Continuity}: the mapping
$(J,\gamma)\longrightarrow \vert J ,\gamma\rangle\; $ is continuous in some appropriate
topology.

\item {\em Resolution of the identity}:
$\int
|J,\gamma\rangle ,\langle J,\gamma|\;dm(J,\gamma)= I$, where $I$ is the
identity in the Hilbert space and $dm$ is some appropriate measure;

\item {\em Temporal stability}:
$e^{-iHt}|J ,\gamma\rangle=|J ,\gamma+\omega t\rangle$, for some
constant $\omega$;

\item {\em Action identity}: $\langle
J,\gamma|H|J ,\gamma\rangle=\omega J$.
\end{itemize}

Their construction, which we shall review below,  works if $H$ has no degenerate eigenstates and,
furthermore, if the lowest eigenvalue is exactly zero. This second
requirement can always be imposed  for reasonable physical systems,
since all physically relevant Hamiltonians $H$ must be bounded from
below, in order to  admit a ground state. This means that there
exists a lowest eigenvalue $E_{min}>-\infty$, so that we can define
a new Hamiltonian, $\widetilde H=H- E_{min}I$, whose lowest
eigenvalue is clearly zero. Furthermore $H$ and $\widetilde H$ have
{\em exactly} the same dynamical content, since they obey
the same commutation relations with all the observables of the system. For
such a Hamiltonian, in the Gazeau-Klauder scheme, one  writes the eigenvalues as
$E_n=\omega\epsilon_n$ by introducing a sequence of dimensionless
quantities $\{\epsilon_n\}$ ordered as follows:
$0=\epsilon_0<\epsilon_1<\epsilon_2<\ldots\;$. Then, the Gazeau-Klauder coherent states are
defined as
\be
|J ,\gamma\rangle:=\mathcal N(J)^{-\frac 12}\sum_{k=0}^\infty\,\frac{J^{n/2}e^{-i\epsilon_n
\gamma}}{\sqrt{\rho_n}}|n\rangle\label{31}
\en
where $\mathcal N$ is a normalization factor, which turns out to be dependent on
$J$ only,  the $|n\rangle$ are the eigenstates of $H$
and the $\rho_n$ are positive numbers, which are fixed by the requirement of the action identity to be
$\rho_n=\epsilon_1\epsilon_2\cdots\epsilon_n$.

  In the rest of this section we recapitulate the Gazeau-Klauder construction, with the
aim of putting
the discussion in a somewhat more general context, which will also enable us to extend the construction
to include vector coherent states and to cases where each energy  level is $(a)$ finitely degenerate
and $(b)$ infinitely degenerate. The essential mathematical ingredient in the construction is a
reproducing kernel Hilbert space.  Although this concept is a familiar one, both in the physical and
the mathematical literature, we summarize below some essential features, putting them in the context
of the present discussion.

\subsection{Some generalities}\label{subsec:somegen}

    Recall that a reproducing kernel Hilbert space (see, for example,
\cite{AAG-book,Aron,Mesch} for detailed discussions)
$\h_{\text{ker}}$, consists of functions $f: X
\longrightarrow \mathbb C$ on some topological space $X$, with the
property that, for all $x \in X$, the evaluation map $E_x :
\h_{\text{ker}} \longrightarrow \mathbb C , \;\; E_x (f) = f(x)$,
is continuous. Such a space may or may not be an $L^2$-space or a
subspace of an $L^2$-space and its scalar product, which we denote
by $\langle\cdot\mid\cdot\rangle_\text{ker}$, may be given in more
general ways. (Although the space $\h_{\text{ker}}$ could be
finite or infinite dimensional, we shall only be interested in the
infinite dimensional case here.) The continuity of the evaluation
map implies that for each $x \in X$, there exists a vector $\xi_x
\in \h_{\text{ker}}$ such that \be
  f(x) = \langle \xi_x \vert f\rangle_\text{ker}\; , \qquad \text{for any } \;\; f \in \h_{\text{ker}}\; .
\label{eval-map}
\end{equation}
The vectors $\xi_x , \; x \in X$, are total in $\h_{\text{ker}}$ (i.e., their
linear span is dense in the space), as can be easily seen. Furthermore, they can  be used to
define the reproducing kernel, $K: X \times X \longrightarrow \mathbb C$,  for this space:
\be
  K(y,x) : = \langle \xi_y \vert \xi_x \rangle_\text{ker} = \xi_x (y)\; ,
\label{repkerdef}
\end{equation}
the second equality following from (\ref{eval-map}). If now $\{\Psi_n\}_{n=o}^\infty$ is an orthonormal
basis  of $\h_{\text{ker}}$, then writing
$$ \xi_x = \sum_{n=0}^\infty \lambda_n (x)\Psi_n\;,  \qquad \lambda_n (x ) = \langle \Psi_n\vert\xi_x\rangle_{\text{ker}}
= \overline{\Psi_n (x)}\;, $$ and taking account of
(\ref{repkerdef}), we get \be K(x,y) = \sum_{n=0}^\infty\Psi_n (x)
\overline{\Psi_n (y)}\; . \label{refkerdef2}
\end{equation}
It ought to be noted that the above equation is true for any orthonormal basis, so that
the kernel $K(x,y)$ is independent of the basis chosen to express it.  An equivalent condition
for the existence of a reproducing kernel is that there be an orthonormal basis for which,
\be
 \sum_{n = 0}^\infty \vert\Psi_n (x)\vert^2   < \infty\; , \qquad \text{for all\;\;} x \in X\; .
\label{repkercond}
\end{equation}

  If we {\em symbolically} write the scalar product of $\h_{\text{ker}}$ as
$$ \langle f \vert g \rangle_\text{ker} = \int_X \overline {f(x)} g(x)\; d\mu (x) \; , $$
then using (\ref{eval-map}) and (\ref{repkerdef})  we may also write
$$ \langle \xi_x \vert \xi_y \rangle_\text{ker}  = \int_X \overline{\xi_x (z)} \xi_y (z)\; d\mu (z) =
  \int_X \langle \xi_x \vert\xi_z\rangle_\text{ker}\; \langle \xi_z \vert\xi_y\rangle_\text{ker}\; d\mu (z)\;. $$
Referring again to (\ref{repkerdef}) and noting that the vectors $\xi_x$ are total in
$\h_{\text{ker}}$, the above equation may be re-expressed either as
\be
  K(x,y) = \int_X K(x, z )K (z, y)\; d\mu (z) \; ,
\label{abs-rep-cond1}
\end{equation}
or as
\be
  \int_X \vert \xi_z\rangle\langle \xi_z \vert\; d\mu (z) = I_\text{ker}\; ,
\label{abs-rep-cond2}
\end{equation}
where $I_\text{ker}$ is the identity operator on $\h_{\text{ker}}$. Thus, these equations
appear now as the well-known
{\em reproducing property} for the kernel $K(x,y)$ and the {\em resolution of the identity} generated by
the vectors $\xi_x$, respectively.  Once more we emphasize that in general,
equations (\ref{abs-rep-cond1}) and
(\ref{abs-rep-cond2}) only have symbolic meaning. However, if in fact $\h_{\text{ker}}$ is an $L^2$-space
with respect to some real measure $d\mu$ on $X$ (or a subspace of such a space), then the above equations
do make literal sense. In view of equations (\ref{abs-rep-cond1}) and (\ref{abs-rep-cond2}), we may call the
vectors $\xi_x$ the {\em coherent states} defined by the kernel $K(x,y)$ and they in fact characterize the reproducing
kernel Hilbert space $\h_{\text{ker}}$. However, since $\Vert\xi_x\Vert^2 = K(x,x)$, these states
are generally not normalized. If $K(x,x) \neq 0$, we may define the normalized vectors $\zeta_x =
[K(x,x)]^{-\frac 12}\xi_x$, for which we would have the ``resolution of the identity''
$$
  \int_X \vert \zeta_z\rangle\langle \zeta_z \vert\;K(x,x)\; d\mu (z) = I_\text{ker}\; . $$

   Coherent states, of all types appearing in the physical literature, can be built by simply transporting
the above structure to some other appropriate Hilbert space by a basis change. To see this,
let $\h$ be an abstract (separable, complex)
Hilbert space and $\{\phi_n\}_{n=0}^\infty$ an orthonormal basis of it. Define the unitary map, $V:
\h_\text{ker} \longrightarrow \h$ by $V\vert \Psi_n \rangle = \vert \phi_n \rangle , \;\; n =0,1,2, \ldots\; .$
Then the  vectors
\be
  \vert \eta_x\rangle  := V\vert\xi_x\rangle = \sum_{n=0}^\infty \overline{\Psi_n (x)}\vert \phi_n\rangle\; ,
\label{absCS}
\end{equation}
define (non-normalized) coherent states on $\h$. They are associated to the same reproducing kernel as the
$\xi_x$ since,
$$ K(x,y) = \langle \eta_x \vert \eta_y \rangle_\h = \langle \xi_x \vert \xi_y\rangle_\text{ker} $$
and satisfy a ``resolution of the identity'' similar to (\ref{abs-rep-cond2}):
$$ \int_X \vert \eta_z\rangle\langle \eta_z \vert\; d\mu (z) = I_\h \; , $$
where again, this equation is to be generally interpreted in the sense of (\ref{abs-rep-cond1}).
Furthermore, for arbitrary $\phi \in \h$, the function $f(x) = \langle \eta_x \vert \phi\rangle_\h$ defines
a vector in $\h_\text{ker}$ and it is easy to see that the inverse of the isometry $V$ is given by this
relation, i.e.,
$V^{-1}\phi (x) =   \langle \eta_x \vert \phi\rangle_\h$.  Usually,
in the physical literature one works with the normalized vectors
\be
  \vert x \rangle = [K(x,x)]^{-\frac 12}\vert\eta_x\rangle =
     [K(x,x)]^{-\frac 12}\sum_{n=0}^\infty \overline{\Psi_n (x)}\vert \phi_n\rangle\; ,
\label{phys-cs}
\end{equation}
It will later  become apparent that the above coherent states coincide  with $\vert J , \gamma\rangle$ in (\ref{31})
upon identifying $\Psi_n (x)$ with $J^{\frac n2} e^{i\epsilon_n\gamma}/\sqrt
{\rho_n}$, $\;\;\rho_n$ with $\epsilon_1 \epsilon_2\cdots \epsilon_n = \epsilon_n! $
and $\mathcal N (J)$ with $K(x,x)$.

   To summarize the preceding discussion, coherent states are linear superpositions of the elements of a
basis in a Hilbert space, the components in the expansion being the values  taken at a point by a set of
vectors forming a basis  in a reproducing kernel Hilbert space. Alternatively, referring to (\ref{repkercond}),
we may identify the reproducing kernel Hilbert space $\h_\text{ker}$ with a subspace of $\ell^2$ generated by
the infinite sequences, $\{\Psi_0 (x),  \Psi_1 (x), \Psi_2 (x), \ldots , \Psi_n (x), \ldots \},\;\; x\in X$.
An associated family of coherent states is then simply given by the vectors,
$\{\overline{\Psi_0 (x)},  \overline{\Psi_1 (x)}, \overline{\Psi_2 (x)}, \ldots , \overline{\Psi_n (x)},
\ldots \},\;\; x\in X$, in this subspace. To see that this way of looking at coherent states does indeed include
all the standard types of coherent states, let us assume that we are give a family of coherent states,
$\vert\lambda \rangle,\;\; \lambda \in \Lambda$, on some Hilbert space $\kk$. The parameter space $\Lambda$ is
assumed to be a topological space. Being coherent states means that the vectors either satisfy a resolution of
the identity,
$$ \int_\Lambda \vert\lambda\rangle\langle\lambda\vert\;dw(\lambda) = I_\kk\; , $$
with respect to some measure $dw$ defined on $\Lambda$, or else that the mapping $\phi \longrightarrow f$, with
$f(\lambda ) = \langle\lambda\vert\phi\rangle$, where $\phi$ runs through $\kk$, is an isometry between $\kk$ and a
reproducing kernel Hilbert space $\kk_\text{ker}$ of functions on $\Lambda$. (In fact the first case implies
the second.) In either case,
if we choose an orthonormal basis $\{\phi_n\}_{n=0}^\infty$ in $\kk$ and expand the coherent states in this basis,
$$ \vert\lambda\rangle = \sum_{n=0}^\infty \overline{f_n (\lambda )}\vert\phi_n\rangle,  \qquad f_n (\lambda )
  = \langle \lambda\vert\phi_n\rangle\; , $$
then the functions $f_n $ are easily seen to form a basis for the  Hilbert space $\kk_\text{ker}$ with
reproducing kernel $K(\lambda , \lambda') = \langle\lambda\vert\lambda'\rangle$.

The above considerations can also be generalized to the case where $\h_\text{ker}$ is a space of
vector valued functions and the kernel $K(x,y)$ is matrix valued, yielding vector coherent
states (see \cite{AEG,sanali}).

\subsection{The Gazeau-Klauder situation}\label{subsec-gazklsit}

   In the light of the preceding discussion, in order to  develop a systematic method for generating
coherent states and vector coherent states of the
Gazeau-Klauder type, we begin by defining  a Hilbert space, $\h_\text{ns}$, of functions
$f: \mathbb R \longrightarrow \mathbb C$, which is complete with respect to the scalar product
\be
  \langle f \mid g\rangle_\text{ns} = \lim_{T\rightarrow \infty} \frac 1{2T}\int_{-T}^T
    \overline{f(\gamma )}g(\gamma )\; d\gamma \; .
\label{ns-scalprod}
\end{equation}
The vectors $f_x , \;\; x \in \mathbb R$,
\be
  f_x (\gamma ) = e^{ix\gamma } \; ,
\label{unitvects}
\end{equation}
are of unit norm and for any two distinct numbers $x, x'$, the corresponding vectors $f_x$
and $f_{x'}$ are orthogonal. This also means that the space $\h_\text{ns}$ is non-separable. Although
this space is not an $L^2$-space, by abuse of notation we shall still symbolically write the scalar product as
\be
  \langle f \mid g\rangle_\text{ns}
   =  \lim_{T\rightarrow \infty} \frac 1{2T}\int_{-T}^T \overline{f(\gamma )}g(\gamma )\; d\gamma
   := \int_{\mathbb R}\overline{f(\gamma )}g(\gamma )\; d\mu (\gamma )\; .
\label{ns-scalprod2}
\end{equation}
If $\{\epsilon_n\}_{n=0}^\infty$ is a sequence of numbers in $\mathbb R$ (we assume that
$\epsilon_n \neq \epsilon_m$
if $n\neq m$), then the set of vectors
\be
   f_n(\gamma ) = e^{i\epsilon_n \gamma }\; , \qquad n = 0, 1, 2, \ldots ,
\label{discr-basis}
\end{equation}
forms a countable orthonormal set and hence the closure of their linear span is
a separable subspace of $\h_\text{ns}$. We
denote this subspace by $\h_\text{ang}$ and it is such subspaces of $\h_\text{ns}$ that we shall use
for constructing coherent states. The reason for the subscript will become clear presently. Suppose next, that
the sequence $\{\epsilon_n\}_{n=0}^\infty$ is so chosen that the following conditions are satisfied,
\bee
\item[(1)] $\epsilon_0 = 0$ and the series
 $$  \sum_{n=0}^\infty \frac {J^n}{\epsilon_n !}\; , \qquad J \in \mathbb R^+ \; , \qquad
    \epsilon_n ! = \epsilon_1 \epsilon_2 \epsilon_3 \cdots \epsilon_n, \qquad \epsilon_0 ! = 1, $$
has a radius of convergence $L > 0$.
\item[(2)] There exists a measure $d\nu$ on $\mathbb R^+$ which solves the moment problem
$$ \int_0^L J^n \; d\nu (J) = \epsilon_n !\; , \qquad \int_0^L d\nu (J) = 1\; . $$
\ene

Then the  vectors $r_n , \;n = 0, 1,2, \dots , $ in $L^2 ( (0, L), d\nu )$ defined by
\be
   r_n (J) = \frac {J^{\frac n2}}{\sqrt{\epsilon_n !}}\; ,
\label{ac-orthon}
\end{equation}
are of unit norm and span the space. Thus the vectors
\be
\Psi_n = r_n \otimes f_n , \quad  \Psi_n (J,\gamma) =
\frac {J^{\frac n2}e^{i\epsilon_n \gamma }}{\sqrt{\epsilon_n !}}\; ,
\qquad n=0, 1, 2,3,  \ldots ,
\label{ac-ang-bas}
\end{equation}
form an {\em orthonormal basis}  in the Hilbert space $\h_\text{ac-ang} = L^2 ( (0, L), d\nu )\otimes
\h_\text{ang}$. Since the vectors $\Psi_n$ satisfy the condition (analogous to (\ref{repkercond})),
\be
  \sum_{n=0}^\infty \vert \Psi_n (J, \gamma )\vert^2 = \sum_{n=0}^\infty
  \frac {J^n}{\epsilon_n !}:= \mathcal N (J)
   < \infty \; ,
\label{repker-cond}
\end{equation}
for all $(J,\gamma ) \in (0,L)\times \mathbb R^+$, the space $\h_\text{ac-ang}$ is a reproducing
kernel Hilbert space with kernel
\be
  K(J, \gamma ;\; J', \gamma' ) = \sum_{n=0}^\infty \Psi_n (J, \gamma ) \overline{\Psi_n (J' , \gamma' )}
       = \sum_{n=0}^\infty \frac {(JJ')^{\frac n2}\; e^{i\epsilon_n (\gamma - \gamma')}}{\epsilon_n !} \; .
\label{ac-ang-repker}
\end{equation}
By (\ref{repkerdef}), the  (non-normalized) coherent states, $\xi_{J,\gamma}$,
defined on $\h_\text{ac-ang}$ and associated to this kernel are then:
\be
   \xi_{J, \gamma}(J'\gamma')  = K (J' \gamma' ; \; J, \gamma ) = \langle \xi_{J', \gamma'} \mid
      \xi_{J, \gamma} \rangle_\text{ac-ang}\; ,
\label{ac-ang-CS}
\end{equation}
while for any $\Psi \in \h_\text{ac-ang}$, we have the relation,
  $$ \langle \xi_{J,\gamma} \mid \Psi \rangle_\text{ac-ang} = \Psi (J, \gamma )\; .$$
Adopting the notation of (\ref{ns-scalprod2}), we may also symbolically write a resolution of the
identity as,
\be
  \int_0^L\!\left[\int_{-\infty}^\infty \vert \xi_{J,\gamma}\rangle\langle \xi_{J,\gamma}\vert
     \; d\mu (\gamma )\right]\; d\nu (J) = I_\text{ac-ang}\; ,
\label{ac-angle-resolid}
\end{equation}
where $I_\text{ac-ang}$ denotes the identity in $\h_\text{ac-ang}$. The above equation is to be
understood in the sense that for arbitrary $\Phi , \Psi \in \h_\text{ac-ang}$,
\begin{align*}
 & \int_0^L\!\left[\int_{-\infty}^\infty \langle \Psi \mid
   \xi_{J,\gamma}\rangle\langle \xi_{J,\gamma}\vert
     \Phi\rangle\; d\mu (\gamma )\right]\; d\nu (J) \\
     & \qquad = \int_0^L\!\left[\lim_{T\rightarrow \infty}\frac 1{2T}
     \int_{-T}^T \overline{\Psi (J, \gamma )}
     \Phi (J, \gamma )\; d\gamma\right]\; d\nu (J)
      =  \langle\Psi\mid \Phi\rangle\; .
\end{align*}
In the Gazeau-Klauder construction of coherent states, related to Hamiltonians with discrete spectra,
one assumes that the Hamiltonian is given on some abstract Hilbert space $\h$ in the
orthonormal basis $\{\phi_n\}_{n=0}^\infty$ by
\be
   H = \omega\sum_{n=0}^\infty \epsilon_n \vert \phi_n \rangle\langle\phi_n\vert\; , \qquad \epsilon_0 = 0\; ,
\label{gaz-klau-ham}
\end{equation}
where $\omega$ is a constant with the dimensions of energy (we take $\hbar = 1$).
The variable $J$ is then generally identified with the classical action and $\gamma$ with the conjugate
angle. It is this identification that prompted our choice of the subscripts for the Hilbert spaces
$\h_\text{ang}$ and $\h_\text{ac-ang}$.

    Following (\ref{absCS}) we can now construct the {\em non-normalized} Gazeau-Klauder type coherent
states in $\h$ using the vectors (\ref{ac-ang-bas}),
\be
  \vert \eta_{J, \gamma }\rangle  = \sum_{n=0}^\infty\overline{\Psi_n (J, \gamma )}\;\vert \phi_n \rangle
     =  \sum_{n=0}^\infty \frac {J^{\frac n2}\;e^{-i\epsilon_n\gamma}}{\sqrt{\epsilon_n !}}\vert \phi_n \rangle\; .
\label{GKCS1}
\end{equation}
Once again, the map
$$W:\h \longrightarrow \h_\text{ac-ang}, \qquad  (W\phi)(J, \gamma )
  = \langle \eta_{J, \gamma }\mid \phi\rangle_\h \; , $$
is  unitary. If instead, we use the normalized vectors,
\be
  \vert J,\gamma\rangle = \mathcal N (J)^{-\frac 12}\vert \eta_{J, \gamma}\rangle\; ,
\label{norm-ac-an-CS}
\end{equation}
with $\mathcal N$ as in (\ref{repker-cond}), the resolution of the identity becomes
\be
  \int_0^L\!\left[\int_{\mathbb R} \vert J,\gamma \rangle\langle J,\gamma \vert
  \;\mathcal N (J)\;d\mu (\gamma )\right]\;d\nu(J) = I_\h \; .
\label{GK-resolid}
\end{equation}
We also have the formal reconstruction formula,
\be
  \vert\phi\rangle = \int_0^L\!\left[\int_{\mathbb R} \Phi (J, \gamma )\vert J,\gamma \rangle
  \;\mathcal N (J)\;d\mu (\gamma )\right]\;d\nu(J)\; , \qquad \Phi (J, \gamma ) =
  \langle J, \gamma \mid \phi\rangle_\h\; ,
\label{rec-form}
\end{equation}
which easily follows from (\ref{GK-resolid}).

   The Gazeau-Klauder coherent states are characterized by the {\em temporal stability} property,
\be
   e^{-iHt}\vert J, \gamma \rangle = \vert J, \gamma + \omega t \rangle\; ,
\label{temp-stab}
\end{equation}
and the {\em action identity},
\be
  \langle J, \gamma \mid H\mid J, \gamma \rangle_\h = \omega J\; .
\label{gk-ac-iden}
\end{equation}
If for a given Hamiltonian, $\epsilon_0 \neq 0$, we will  work with the new Hamiltonian
$H' = H - \omega\epsilon_0 I_\h$, and use $\varepsilon_n = \epsilon_n - \epsilon_0$ to construct coherent
states. Note that this amounts to simply shifting all the energy levels by a constant so as to
bring the ground state energy to zero and moreover, the new Hamiltonian commutes with the old Hamiltonian.
In this case,
\bea
 e^{-iHt}\vert J, \gamma \rangle & = & e^{-iH' t}\;e^{-i\omega \epsilon_0 t}\vert J, \gamma \rangle
                                = e^{-i\omega \epsilon_0 t}\vert J, \gamma + \omega t
                                \rangle\nonumber\\
 \langle J, \gamma \mid H \mid J , \gamma \rangle_\h & = &
      \langle J, \gamma \mid H' + \omega\epsilon_0 \mid J , \gamma \rangle_\h = J + \omega\epsilon_0\; .
 \label{mod-conds}
\ena

\section{Vector coherent states of the Gazeau-Klauder type}\label{GK-VCS}
    Suppose now that the Hamiltonian $H$ (acting on the Hilbert space $\h$) has a discrete positive
spectrum and that the eigenvectors $\phi_{jk}, \; \; j=1,2,3,
\ldots , N < \infty, \;\; k = 0,1,2,3, \ldots , \infty$, can be
grouped into $N$ families, each containing an infinite number of
vectors. (Such a situation could arise, for example, through the
lifting of an $N$-fold degeneracy in the energy spectrum, by an
interaction. Therefore $k$ labels the {\em main} energy levels
while $j$ labels the {\em sublevels} generated by, e.g., a small
perturbation.) Furthermore, assume that the corresponding
eigenvalues $E_{jk} = \omega \epsilon_{jk}$ satisfy $\epsilon_{j0}
= 0, \;\; j = 1,2,3, \ldots , N$, and for any  $j$, $\epsilon_{jk}
\neq \epsilon_{j'\ell}$ if $k \neq \ell$ and $\forall \,j,j'$.
Denote by $\h_j$ the subspace of $\h$ spanned by the vectors
$\phi_{jk}, \; k =0,1,2, \ldots , \infty ,$ and by $\mathbb P_j$
the projection operator onto this subspace. Then $H = \oplus_{j =
1}^N H_j$, with $H_j = \omega\sum_{k=0}^\infty
\epsilon_{jk}\vert\phi_{jk}\rangle\langle\phi_{jk}\vert$, which
leaves $\h_j$ stable. We will give an example of such a
decomposition in the first application below. In $\h_j$ we define
the coherent states, \be
  \vert J_j , \gamma_j\rangle = \mathcal N (J_j)^{-\frac 12} \sum_{k=0}^\infty
     \frac {J_j^{\frac k2}\; e^{-i\epsilon_{jk}\gamma_j}}{\sqrt{\epsilon_{j1}\epsilon_{j2}\ldots
     \epsilon_{jk}}}\;
     \vert\phi_{jk}\rangle\; .
\label{par-vcs1}
\end{equation}
Here $-\infty < \gamma_j < \infty$ and $0\leq J_j  < L_j = \lim_{k\rightarrow \infty} \epsilon_{jk}$,
and we assume that $L_j  > 0$. The normalization factor $\mathcal N (J_j )$ is chosen so that
\be
\langle J_j , \gamma_j \mid J_k , \gamma_k \rangle = \delta_{jk}\;,
\label{par-vcs2}
\end{equation}
These states also satisfy
\be
  e^{-iH_j t}\vert J_j , \gamma_j\rangle = \vert J_j , \gamma_j+ \omega t \rangle\; , \qquad
  \langle J_j , \gamma_j \mid H_k \mid J_k , \gamma_k \rangle = \omega J_j \delta_{jk}\; ,
\label{par-vcs3}
\end{equation}
and the ``partial resolution of the identity'':
\be
  \int_0^{L_j}\!\left[\int_{\mathbb R} \vert J_j ,\gamma_j \rangle\langle J_j ,\gamma_j  \vert
  \;\mathcal N (J_j)\;d\mu (\gamma_j )\right]\;d\nu_j(J_j) = \mathbb P_j\; ,
\label{par-resolid}
\end{equation}
where $d\mu$ is as in (\ref{ns-scalprod2}) and the measure $d\nu_j (J_j )$ is
defined through the moment problem
\be
 \int_0^{L_j} J^n \; d\nu_j (J) = \epsilon_{j1}\epsilon_{j2}\ldots \epsilon_{jn} \; ,
 \qquad \int_0^{L_j} d\nu_j (J) = 1\; .
\label{par-mom-prob}
\end{equation}

  Next, introducing  the diagonal matrices,
\bea
  \bJ =\text{diag}\;(J_1 , J_2 , \ldots , J_N )\; , &\qquad& \bvarepsilon_k =
  \text{diag}\; (\epsilon_{1k}, \epsilon_{2k}, \ldots , \epsilon_{Nk})\; ,\nonumber\\
  \bgamma = \text{diag}\; (\gamma_1 , \gamma_2 , \ldots , \gamma_N )\; , &\qquad&
  \bvarepsilon_k ! = \bvarepsilon_1\bvarepsilon_2\ldots\bvarepsilon_k \; ,
\label{mat-var1}
\ena
and the vectors
\be
\vert \bPhi_k ; \; j\rangle = \begin{pmatrix} 0 \\ \vdots \\ \vert \phi_{jk}\rangle \\\vdots \\ 0
   \end{pmatrix}\; , \quad j=1,2, \ldots , N, \quad k =0, 1,2, \ldots \; ,
\label{mat-var2}
\end{equation}
we may rewrite the vectors (\ref{par-vcs1}) as
\bea
\vert \bJ , \bgamma ;\; j\rangle &  : = &
  \mathcal N (J_j )^{-\frac 12}\sum_{k=0}^\infty \left[\bvarepsilon_k !\right]^{-\frac 12}\;
     \bJ^{\frac k2}\; \exp [-i\bvarepsilon_k\bgamma ]\; \vert\bPhi_k ;\; j\rangle
     \nonumber\\
     & = &
     \begin{pmatrix} 0 \\ \vdots \\ \vert J_j , \gamma_j \rangle \\ \vdots \\ 0 \end{pmatrix}\; .
\label{gk-vcs1}
\ena
We call these states {\em vector coherent states} for the Hamiltonian $H$. Note that, in this representation,
$H$ is a diagonal operator, $H = \text{diag} \;(H_1 , H_2, \ldots , H_N )$, each $H_j$ being
an infinite diagonal matrix with eigenvalues $\omega\epsilon_{jk} , \; k = 0,1,2, \ldots \; .$
\be
  e^{-iHt}\vert \bJ , \bgamma ;\; j\rangle = \vert \bJ , \bgamma + \omega t \mathbf d_j;\; j\rangle\;,
  \qquad \langle \bJ , \bgamma; \; j \mid H\mid \bJ , \bgamma ; \; j\rangle = \omega J_j\; ,
\label{gk-vcs2}
\end{equation}
where $\mathbf d_j$ is the diagonal matrix with one in the $jj$-position and zeroes elsewhere. Furthermore,
we have the resolution of the identity on $\h$:
\be
   \sum_{j=0}^N\int_0^{L_N}\!\!\ldots\int_0^{L_1}\!\left[\int_{\mathbb R^N}
   \vert \bJ ,\bgamma ;\; j\rangle\langle \bJ ,\bgamma ;\; j  \vert
  \;\mathcal N (J_j)\;d\mu (\bgamma)\right]\;
  d\nu(\bJ) = I_\h \; ,
\label{gk-resolid}
\end{equation}
with
$$d\nu (\bJ ) = d\nu (J_1 )\; d\nu (J_2 )\;\ldots\; d\nu (J_N )\; , \qquad
d\mu (\bgamma) = d\mu_1 (\gamma_1 )\; d\mu_2 (\gamma_2 )\; \ldots\; d\mu_N (\gamma_N ). $$

    In view of the fact that (see also (\ref{par-vcs2}))
\be
  \langle \bJ , \bgamma ;\; j\mid \bJ , \bgamma ;\; k\rangle = \delta_{jk} \; ,
\label{vcs-orthog}
\end{equation}
a general vector coherent state for such a system may be written as a linear combination,
$$
  \vert \bJ , \bgamma \rangle = \sum_{j=0}^N c_j \vert \bJ , \bgamma ;\; j\rangle \; .$$
However, such a state would, in general, not be of the Gazeau-Klauder type, unless the levels
$\epsilon_{jk}, \; j=1,2, \ldots , N$, are degenerate for all $k$. Associated to the vector
coherent states (\ref{gk-vcs1}) is the {\em matrix-valued} reproducing kernel,
$\bK (\bJ , \bgamma,\; \bJ' , \bgamma' )$, with matrix elements
\be
\bK (\bJ , \bgamma;\; \bJ' , \bgamma' )_{jk} = \langle \bJ , \bgamma;\; j \mid \bJ' , \bgamma';\; k\rangle\; .
\label{mat-rep-ker1}
\end{equation}
This kernel has the properties,
\begin{align}
& \bK (\bJ , \bgamma;\; \bJ , \bgamma )_{jj}  =
  \Vert\; \vert \bJ , \bgamma ;\; j\rangle\; \Vert^2  > 0\; ,
  \qquad
\bK (\bJ , \bgamma;\; \bJ' , \bgamma' )_{jk}  =  \overline{\bK (\bJ' , \bgamma';\; \bJ , \bgamma )_{kj}}\; ,
\nonumber\\[3mm]
&  \sum_{\ell =0}^N\int_0^{L_N}\!\!\ldots\int_0^{L_1}\!
     \left[\int_{\mathbb R^N}
  \bK (\bJ , \bgamma;\; \bJ'' , \bgamma'' )_{j\ell }\;\bK (\bJ'' , \bgamma'';\; \bJ' , \bgamma' )_{\ell k}
  \;\mathcal N (J_\ell)\;d\mu (\bgamma'')\right]\; d\nu(\bJ'')\nonumber\\
  & \qquad\qquad = \bK (\bJ , \bgamma;\; \bJ' , \bgamma' )_{jk} \; .
\label{mat-repker2}
\end{align}

\subsection{Some examples}
Let us consider a model described by the following Hamiltonian,
\be H=\omega a^\dagger a+\epsilon_1c_1^\dagger
c_1+\epsilon_2c_2^\dagger c_2+ (g_1c_1^\dagger c_1+g_2c_2^\dagger
c_2)(a+a^\dagger)\label{11} \en where the following commutation
rules hold:
\be
[a,a^\dagger]=\{c_1,c_1^\dagger\}=\{c_2,c_2^\dagger\}=I,
\label{12}\en
and
\be
[a^\sharp,c_i^\sharp]=\{c_1,c_1\}=\{c_2,c_2\}=0, \label{13}\en
where $a^\sharp$ stands for $a$ or $a^\dagger$, $[A,B]=AB-BA$ and
$\{A,B\}=AB+BA$. This model, which describes an interaction between
a single mode, ($a, a^\dagger$), of the radiation field with two Fermi
type modes, has been analyzed quite recently in \cite{simon}.

A convenient feature of the above hamiltonian is that its spectrum can
be obtained explicitly, as well as its eigenvectors. In fact, considering the
{\em fermionic} part, it is clear that all the eigenstates
of $H$ must be of the following form:
\be
\Phi=\varphi\otimes\Psi_{kl}, \mbox{ where }
\Psi_{kl}=\left(c_1^\dagger\right)^k\left(c_2^\dagger\right)^l\Psi_{00},
\label{14}
\en
with $k,l=0,1$, and where $\Psi_{0,0}$ is the fermionic vacuum:
$c_j\Psi_{00}=0$, for $j=1,2$. The vector
$\varphi$ has still to be determined, but it is clear that it cannot, in general,  be
proportional to $\left(a^\dagger\right)^n\varphi_0$,  where
$a\varphi_0=0$, since the interaction part of $H$ is
not diagonal on these vectors. However it is a rather
simple exercise to check that
\bea
 \left\{
    \begin{array}{ll}
        & H \left(\varphi\otimes\Psi_{00}\right)=
        \omega a^\dagger a\left(\varphi\otimes\Psi_{00}\right) \\
       & H \left(\varphi\otimes\Psi_{10}\right)=
        \left(\omega a^\dagger a+\epsilon_1+g_1(a+a^\dagger)\right)
        \left(\varphi\otimes\Psi_{10}\right) \\
        & H \left(\varphi\otimes\Psi_{01}\right)=
        \left(\omega a^\dagger a+\epsilon_2+g_2(a+a^\dagger)\right)
        \left(\varphi\otimes\Psi_{01}\right) \\
 & H \left(\varphi\otimes\Psi_{11}\right)=
        \left(\omega a^\dagger a+\epsilon_1+\epsilon_2+(g_1+g_2)(a+a^\dagger)\right)
        \left(\varphi\otimes\Psi_{11}\right) \\
              \end{array}
        \right.
\label{15}
\ena
To proceed further, we observe that in each of the four cases above, $\phi$ is an eigenvector
of an self-adjoint operator of the type,
\be
  B_{k l} = \omega A^\dag_{kl} A_{kl} + \big( \varepsilon_{kl} - \frac {g_{kl}^2}\omega\big)I\; ,
\qquad A_{kl} = a + \frac {g_{kl}}\omega\; , \qquad [A_{kl} , A_{kl}^\dag ] = 1 \; ,
\quad k,l = 0,1\; ,
\label{aux-hamilt}
\end{equation}
where,
$$ \varepsilon_{kl} = l\epsilon_1 + k\epsilon_2\; , \qquad g_{kl} = lg_1 + kg_2\; , \qquad l,k = 0,1\; .
$$
We know, however, that
$$ A_{kl} = \exp\!\big[i\sqrt{2}\;\frac {g_{kl}}{\omega}\; P\big]\; a\;
\exp\!\big[- i\sqrt{2}\;\frac {g_{kl}}{\omega}\; P\big]\;,  \qquad \text{where} \quad
P = \frac {a - a^\dag }{i\sqrt{2}}\; .
$$
Thus, the eigenvectors of $B_{kl}$ are,
\be
 \vert \Phi_n^{kl}\rangle =  \exp\!\big[i\sqrt{2}\;\frac {g_{kl}}{\omega}\; P\big]\vert n\rangle
 = \frac {(A_{kl}^\dag )^n}{\sqrt{n!}} \vert \Phi_0^{kl}\rangle \; ,
\label{aux-hamilt3}
\end{equation}
where $ \vert n \rangle = \displaystyle{\frac {a^n}{\sqrt{n!}}}\vert 0 \rangle$ are the eigenvectors of
the usual number operator $N = a^\dag a$.

    The diagonalization of $H$ is now complete. Our results can be summarized as follows:

\vspace{2mm}

\noindent{\em eigenstates of $H$}:
$\{\varphi_{n}^{kl}:=\Phi_{n}^{kl}\otimes\Psi_{kl}, \text{
where } n = 0,1,2, \ldots\; ,  \, \text{ and } k,l=0,1\}$

\vspace{2mm}

\noindent{\em eigenvalues of $H$}: $\{E_{n}^{kl}, \text{ with }
n = 0,1,2, \ldots \; , \, \text{ and } k,l=0,1\}$,

\vspace{2mm}

\noindent where the relevant quantities are shown in the following
table:

\vspace{2mm}

\noindent\begin{footnotesize}
 \begin{tabular}{||c||c|c|c|c|c||}
\hline
\hline $k, l$           &$E_{n}^{kl} = $
&$\Psi_{kl} = $ & $\Phi_{n}^{kl} = $ & where  & and\\
\hline \hline$0, 0$           &$\omega n$
&$\Psi_{00}$ &$\frac{(a^\dagger)^n}{\sqrt{n!}}\Phi_{0}^{00}$ & $a\Phi_{0}^{00}=0$  & \\
\hline $1, 0$           &$\omega
n+\epsilon_1-\frac{g_1^2}{\omega}$ & $c_1^\dagger\Psi_{00}$
& $\frac{(A_{10}^\dagger)^n}{\sqrt{n!}}\Phi_{0}^{10}$
& $A_{10}\Phi_{0}^{10}=0$  & $A_{10}=a+\frac{g_1}{\omega}$\\
\hline  $0, 1$           &$\omega
n+\epsilon_2-\frac{g_2^2}{\omega}$ &$ c_2^\dagger\Psi_{00}$
&$\frac{(A_{01}^\dagger)^n}{\sqrt{n!}}\Phi_{0}^{01}$
& $A_{01}\Phi_{0}^{01}=0$  & $A_{01}=a+\frac{g_2}{\omega}$\\
\hline $1, 1$           &$\omega
n+\epsilon_1+\epsilon_2-\frac{(g_1+g_2)^2}{\omega}$ &$c_1^\dagger
c_2^\dagger\Psi_{00}$
&$\frac{(A_{11}^\dagger)^n}{\sqrt{n!}}\Phi_{0}^{11}$
& $A_{11}\Phi_{0}^{11}=0$  & $A_{11}=a+\frac{g_1+g_2}{\omega}$\\
\hline\hline
\end{tabular}
\end{footnotesize}

\vspace{3mm}

\noindent From (\ref{aux-hamilt3}) it is also clear that the vectors $\Phi_0^{kl}$ are just the well
known canonical coherent states $\vert z \rangle$, with $z = -\displaystyle{\frac {g_{kl}}\omega}$.
Thus, in the position space representation these vectors are shifted
Gaussians,
$$
\Phi_{0}^{kl}(x)\simeq e^{-\frac{1}{2}(x+\sqrt{2}g_{kl}^2)},
\qquad k =0,1\; .
$$

  In order to build Gazeau-Klauder type of coherent states for this Hamiltonian, we see now that it
breaks up into four orthogonal parts:
\be
  H = \oplus_{k,l = 0,1} H_{kl}\; ,  \quad \text{where} \quad H_{kl} = \sum_{n=0}^\infty E_n^{kl}
    \vert \varphi_n^{kl}\rangle \langle \varphi_n^{kl}\vert\; .
\label{H-decomp}
\end{equation}
Since the lowest eigenvalue $E_0^{kl}$, for the component Hamiltonian $H_{kl}$ , is
zero only  for $k=l=0$, we work
with $H' = \oplus_{k,l = 0,1} H'_{kl}$, where $H'_{kl} = \sum_{n=0}^\infty (E_n^{kl} - E_0^{kl})
\vert \varphi_n^{kl}\rangle \langle \varphi_n^{kl}\vert$. But $E_n^{kl} - E_0^{kl} = \omega n$.
(Note that $H$ and $H'$ commute.)
Thus, the vector coherent states of the present model are $4$-component vectors, involving the
standard canonical coherent states, $\vert z_{kl}\rangle , \; k,l = 0,1 , \;\; z_{kl} \in \mathbb C$,
built on the bosonic vacuum state $\Phi_0^{kl}$.
Thus, introducing the diagonal matrix $\mathfrak Z = \text{diag}(z_{00}, z_{10}, z_{01}, z_{11})$,
we can write the vectors (\ref{gk-vcs1}) for the present case as
\be
\vert \mathfrak Z ; \; kl \rangle = \vert z_{kl}\rangle\vert\Psi_{kl}\rangle
  = e^{-\frac {\vert z_{kl}\vert^2}2}\sum_{n=0}^\infty \frac {\mathfrak Z^n}{\sqrt{n!}}
     \vert\Psi_{kl}\rangle\vert\Phi_n^{kl}\rangle\; , \quad j,k = 1,2,
\label{ex1-gk-vcs}
\end{equation}
where in the present representation, the vectors $\Psi_{kl}$ form the canonical basis of $\mathbb C^4$:
$$
  \Psi_{00} = \begin{pmatrix} 1\\0\\0\\0\end{pmatrix}, \quad
  \Psi_{10} = \begin{pmatrix} 0\\1\\0\\0\end{pmatrix}, \quad
  \Psi_{01} = \begin{pmatrix} 0\\0\\1\\0\end{pmatrix}, \quad
  \Psi_{11} = \begin{pmatrix} 0\\0\\0\\1\end{pmatrix} . \quad
$$
These then are the Gazeau-Klauder type vector coherent states for the Hamiltonian (\ref{11}).
Equations (\ref{gk-vcs2}) and (\ref{gk-resolid}) have obvious transcriptions for these states.

\vspace{3mm}

   One could also consider the following variant of the Hamiltonian (\ref{11}):
\be
H=\omega a^\dagger a+\epsilon_1c_1^\dagger
c_1+\epsilon_2c_2^\dagger c_2+ \sum_{i,j=1}^2g_{ij}c_i^\dagger
c_j(a+a^\dagger)\label{19}
\en
where the same commutation rules (\ref{12}) and (\ref{13}) are assumed and
$$
g=\left(
\begin{array}{cc}
g_{11} & g_{12}   \\
g_{21} & g_{22}  \\
\end{array}
\right)
$$
is a $2\times 2$ hermitiian matrix, $g={\overline g}^T$. Let V be the
unitary matrix which diagonalizes $g$:
$$
VgV^{-1}=g_d:=\left(
\begin{array}{cc}
g_{1} & 0   \\
0 & g_{2}  \\
\end{array}
\right),$$
so that, defining
$$d=\left(
\begin{array}{c}
d_1   \\
d_2  \\
\end{array}
\right)=Vc=V\left(\begin{array}{c}
c_1   \\
c_2  \\
\end{array}
\right),
$$
and $d^\dagger=c^\dagger V^\dagger=(d_1^\dagger , \; d_2^\dagger)$, the
operators $d_j$ again obey  the same anticommutation relations as the $c_j$. Also,
$\sum_{i,j=1}^2g_{ij}c_i^\dagger c_j=g_1d_1^\dagger
d_1+g_2d_2^\dagger d_2$.  However, if $\epsilon_1\neq\epsilon_2$,
this change of variables would make the free fermionic Hamiltonian
$\epsilon_1c_1^\dagger c_1+\epsilon_2c_2^\dagger c_2$ no longer
diagonal, while if $\epsilon_1=\epsilon_2=\epsilon$ we get
$$H=\omega a^\dagger a+\epsilon d_1^\dagger
d_1+\epsilon d_2^\dagger d_2+ (g_1d_1^\dagger d_1+g_2d_2^\dagger
d_2)(a+a^\dagger),$$
for which the entire analysis  performed above
can be repeated.

\vspace{2mm}

{\bf Remark\,\,} A possible method for describing a non-degenerate
two-level atom (i.e., $\epsilon_1 \neq \epsilon_2$), which is the
one considered in \cite{simon}, can be obtained by adapting the
previous procedure  as follows: we consider a fictitious
three-level atom interacting with the radiation field in the
following way:
$$H=\omega a^\dagger a+\epsilon (c_1^\dagger
c_1+ c_2^\dagger c_2+ c_3^\dagger c_3)+
\sum_{i,j=1}^3g_{ij}c_i^\dagger c_j(a+a^\dagger),$$
where now $\{g_{ij}\}$ is a $3\times 3$ hermitian matrix. We recover a
two-level system by considering a subspace of the complete Hilbert
space spanned by the vectors
$\Psi_{kl}\otimes\Phi_n^{kl}$, where the $\Phi_n^{kl}$ are
constructed by trivially extending the foregoing procedure. Next we take
$\Psi_0^{00}=\Psi_0$ to be the ground state of $c_j$,
$j=1,2,3$ and  set $\Psi_{10}=c_1^\dagger \Psi_0$, $\Psi_{01}=c_2^\dagger
c_3^\dagger\Psi_0$ and $\Psi_{11}=c_1^\dagger c_2^\dagger
c_3^\dagger\Psi_0$. (The interpretation is clear: $\Psi_0$ corresponds
to both  levels of our atom being empty, while $\Psi_{10},\; \Psi_{01}$ and $\Psi_{11}$
correspond respectively to the first, second and both levels
being occupied.)

If it is now possible to ensure that the resulting energy spectrum
$E_n^{kl}, \; n=0,1,2, \ldots , \;\; k,l = 0,1$, has no degeneracies,
we could build Grazeau-Klauder type coherent states for this system.
On the other hand, it is easily verified that degeneracy will be avoided
if the physical constants of the model satisfy the following inequalities:
$$
0<\epsilon_1-\frac{g_1^2}{\omega}<\epsilon_2-\frac{g_2^2}{\omega}<
\epsilon_1+\epsilon_2-\frac{(g_1^2+g_2^2)}{\omega}<\omega
$$
In this case we put $E_0=E_{0}^{00}=0,
E_1=E_{0}^{10}=\epsilon_1-\frac{g_1^2}{\omega},\;
E_2=E_{0}^{01}=\epsilon_2-\frac{g_2^2}{\omega},\;
E_3=E_{0}^{11}=\epsilon_1+\epsilon_2-\frac{(g_1^2+g_2^2)}{\omega},\;
E_4=E_{1}^{00}=\omega$, and so on and write, for the corresponding
eigenstates  $\varphi_0=\varphi_{0}^{00},\;
\varphi_1=\varphi_{0}^{10},\; \varphi_2=\varphi_{0}^{01},\;
\varphi_3=\varphi_{0}^{11}, \;\varphi_4=\varphi_{1}^{00},$ and so on.
Finally, defining  $\epsilon_n=\frac{E_n}{\omega}$, we recover a
sequence of quantities satisfying the inequalities
$0=\epsilon_0<\epsilon_1<\epsilon_2<\ldots\;$, as required in
\cite{GK}. Thus we obtain the coherent states $|J,\gamma\rangle
=N(J)^{-\frac 12}\sum_{n=0}^\infty\frac{J^{n/2}e^{-i\gamma
\epsilon_n}}{\sqrt{\epsilon_n!}}\varphi_n$, with all the
required properties.
\vspace{3mm}


\section{Hamiltonians with degeneracies}\label{ham-degen}

Here we extend the preceding construction  to
the situation in which some (or perhaps all) of the eigenvalues of the
given Hamiltonian have degeneracies. We will consider two situations: first,
where all the degeneracies are finite and second, where they are all
countably infinite. In the first case, we will
show that a natural way to recover all the required properties
of the Gazeau-Klauder type coherent states, such as the
resolution of the identity, temporal stability and the action
identity, among others, is to introduce a
third parameter into the definition of the coherent states, replacing
$|J ,\gamma\rangle$ by $|J,\gamma ,\theta\rangle$. The extension
we are proposing is somewhat different from that suggested in
\cite{fox,klau}, since it only involves one extra parameter. Moreover, as we
will demonstrate, our method can also be adapted to the case of infinite
degeneracies.

\subsection{Finite degeneracies}\label{fin-deg}

Let us now consider a Hamiltonian $H$, the eigenvalues of which
are all discrete with the lowest eigenvalue being again zero.
Assume that the $n$-th  level, $E_n=\omega\epsilon_n$, has a
degeneracy $d(n)$, in general different from 1. We assume $d(n) <
\infty$, for all $n$. Denote by $|n,j\rangle , \;
n=0,1,2,\ldots,\;\; j=1,2,\ldots  , d(n)$, the eigenvectors of the
Hamiltonian $H$ so that $H |n,j\rangle=E_n |n,j\rangle$, with $n$
labelling the level and $j$ counting the degeneracy. As usual we
introduce the dimensionless quantity $\epsilon_n$ and again,
without loss of generality, arrange them in the sequence
$0=\epsilon_0<\epsilon_1<\epsilon_2<\ldots\;$. This means that the
hamiltonian is
$H=\omega\sum_{n=0}^\infty\sum_{j=1}^{d(n)}\epsilon_n
|n,j\rangle\langle n,j|$. We next introduce the parameter
$\theta\in [0,2\pi)$ and define \be
|J,\gamma,\theta\rangle:=\mathcal N(J)^{-\frac
12}\sum_{n=0}^\infty\sum_{j=1}^{d(n)}\,
         \frac{J^{n/2}e^{-i\epsilon_n \gamma}e^{-ij\theta}}{\sqrt{\rho_n}}|n,j\rangle,
\label{32} \en with $J$ and $\gamma$ as before. We now prove that,
for appropriate choice of $\rho_n$, these states satisfy the
following properties, which naturally generalize the analogous
ones stated at the beginning of Section \ref{GK-revis} :

\begin{itemize}
\item {\em Continuity}: if $(J,\gamma,\theta)\rightarrow(J',\gamma',\theta')$ then
$|J,\gamma,\mu\rangle\rightarrow |J',\gamma',\mu'\rangle$;

\item {\em Resolution of the identity}: $\int
|J,\gamma,\theta\rangle\,\langle J,\gamma,\theta \vert\; dm(J,\gamma,\theta )= I$, for some appropriately
chosen measure $dm$;

\item {\em Temporal stability}: $e^{-iHt}|J,\gamma,\theta\rangle=|J,\gamma+\omega
t,\theta\rangle$, for some constant $\omega$;

\item {\em Action identity}: $\langle
J,\gamma,\theta\mid H \mid J,\gamma,\theta\rangle=\omega J$.
\end{itemize}

Indeed, continuity follows automatically from the definition
itself. As for normalization, we observe that
\beano
\langle J,\gamma,\theta \mid J,\gamma,\theta \rangle & = &
\mathcal N (J)^{-1} \sum_{n,m=0}^\infty\sum_{j=1}^{d(n)}
\sum_{l=1}^{d(m)}\,\frac{J^{n/2+m/2}e^{-i(\epsilon_n-\epsilon_m)
\gamma}e^{-i(j-l)\theta}}{\sqrt{\rho_n\rho_m}}\langle m,l|n,j\rangle\\
 & = & \mathcal N(J)^{-1} \sum_{n=0}^\infty\sum_{j=1}^{d(n)}
\,\frac{J^{n}}{\rho_n}=\mathcal N(J)^{-1} \sum_{n=0}^\infty
\,\frac{J^{n}d(n)}{\rho_n},
\enano
from which we conclude that $\langle
J,\gamma,\theta \mid J ,\gamma,\theta\rangle=1$ if and only if
\be \mathcal N(J) =
\sum_{n=0}^\infty \,\frac{J^{n}d(n)}{\rho_n}\; .
\label{33}
\en
Of course, this is a power series in $J$ and we assume that it has a
radius of convergence $L > 0$.

The proof of temporal stability is easy:
\beano
e^{-iHt}\vert J,\gamma,\mu\rangle & = &
e^{-iHt}\mathcal N(J)^{-\frac 12}\sum_{n=0}^\infty\sum_{j=1}^{d(n)}\,\frac{J^{n/2}e^{-i\epsilon_n
\gamma}e^{-ij\mu}}{\sqrt{\rho_n}}|n,j\rangle\\
  & = & \mathcal N(J)^{-\frac 12}\sum_{n=0}^\infty\sum_{j=1}^{d(n)}\,\frac{J^{n/2}e^{-i\epsilon_n
\gamma}e^{-ij\mu}}{\sqrt{\rho_n}}e^{-i\omega\epsilon_nt}|n,j\rangle\\
  & = & |J,
\gamma+\omega t, \mu\rangle\; .
\enano

  In order for the action identity to be satisfied, we need a condition on the $\rho_n$. Since
$\epsilon_0 = 0$, we get
$$
 \langle J, \gamma, \theta \mid H \mid J, \gamma , \theta \rangle = \omega J\;
 \left[\mathcal N(J)^{-1}\sum_{n=1}^\infty
   \frac {\epsilon_n J^{n-1} d(n)}{\rho_n}\right]\; .
$$
Thus, in order for the action identity to hold the expression within the square brackets must equal one.
This can be achieved if we require that
$$ \frac {\epsilon_n d(n)}{\rho_n} = \frac {d(n-1)}{\rho_{n -1}}, \;\; n=1,2,3, \ldots \; ,  $$
for then
$$ \rho_n = \epsilon_n \frac {d(n)}{d(n-1)}\rho_{n-1} = \ldots \; =  \epsilon_n !\frac {d(n)}{d(0)} \rho_0\; ,
\quad \text{by iteration}\; . $$
We choose $\rho_0 = d(0)$ so that
\be
  \rho_n = \epsilon_n !\; d(n),\; \;  n = 0,1,2, \ldots \; ,\quad \text{and}\quad \mathcal N(J)
  = \sum_{n=0}^\infty\frac {J^n}{\epsilon_n !}\; .
\end{equation}
Thus the coherent states (\ref{32}) become
\be
|J,\gamma,\theta\rangle:=\mathcal N(J)^{-\frac 12}\sum_{n=0}^\infty\sum_{j=1}^{d(n)}\,
         \frac{J^{n/2}e^{-i\epsilon_n \gamma}e^{-ij\theta}}{\sqrt{\epsilon_n ! d(n)}}\vert n,j\rangle\; .
\label{degen-cs}
\en

 It remains only to determine the measure $dm$ in order for the resolution of the identity
to be satisfied. Proceeding as in Section \ref{GK-revis}, and assuming that the measure $d\nu$
solves the moment problem
\be
 \int_0^L J^n \; d\nu (J) = \epsilon_n ! \; d(n)\; , \qquad  n =0,1,2, \ldots \; ,
\label{deg-mom-prob}
\end{equation}
we take
\be
dm (J, \gamma , \theta ) = \frac {\mathcal N (J)}{2\pi }\; d\nu (J)\; d\mu (\gamma)\; d\theta\; ,
\label{gen-meas}
\en
where $d\mu$ is the symbolic measure defined in (\ref{ns-scalprod2}). Then, we prove exactly as in
Section \ref{GK-revis}, the identity (see (\ref{GK-resolid}))
\be
  \frac 1{2\pi}\;\int_0^L\!\left\{\int_0^{2\pi}\!\left[\int_{\mathbb R} \vert J,\gamma , \theta \rangle
  \langle J,\gamma , \theta  \vert
  \;\mathcal N (J)\;d\mu (\gamma )\right] d\theta\right\} d\nu(J) = I_\h \; .
\label{GK-degen-resolid}
\end{equation}

{\bf Remark:} If $d(n)=1$ for all $n$, the above
coherent states coincide, apart from an inessential overall
phase $e^{-i\theta}$, with the usual Gazeau-Klauder coherent states (\ref{31}).
However, when the Hamiltonian $H$ has a non-trivial degeneracy, it is
interesting to notice the presence of $d(n)$ in the denominator of the
expression for the coherent states in (\ref{degen-cs}), which implies that
the radius of convergence $L$ depends not only on the eigenvalues of the Hamiltonian
but also on their degeneracies. Similarly, the measure $d\nu$, solving the moment problem
(\ref{deg-mom-prob}) and appearing in the resolution of the identity, depends on the
degeneracy.

\vspace{2mm}

{\bf Example 1:} Consider the following simple example, consisting of a single
boson and a single  fermion: $ H=\omega(a^\dagger
a+c^\dagger c),$ where $[a,a^\dagger]=\{c,c^\dagger\}= I$ and
$[a^\sharp,c^\sharp]=0$, $x^\sharp$ being $x$ or $x^\dagger$.
Introducing the vacuum  $\Phi_0$ of $a$, and  $\Psi_0$ of $c$ and taking, as usual
$\Phi_n=\displaystyle{\frac{(a^\dagger)^n}{\sqrt{n!}}}\Phi_0 ,\;\;
n=0,1,2,\ldots\; , $ and $\Psi_j=(c^\dagger)^j\Psi_0 , \;\; j=0,1$, we can
write the eigenvectors of $H$ as $\varphi_{n,j}=\Phi_{0}\otimes\Psi_0$ if $n=j=0$, and
$\varphi_{n,j}=\Phi_{n-j}\otimes\Psi_j$, if $n=1,2,3,\ldots \; , $ and
$j=0,1$. The corresponding eigenvalues are $E_{n,j}=n\omega$, so
that they turn out to be degenerate in $j$. In particular we have
$d(0)=1$ and $d(n)=2$ for all $n\geq 1$. The normalization can be
computed using  (\ref{33}), and we get
$$\mathcal N(J) =
\sum_{n=0}^\infty \frac{J^{n}d(n)}{\rho_n}= 1+ 2\sum_{n=1}^\infty
\frac{J^{n}}{\rho_n} = 1+ \sum_{n=1}^\infty \frac{J^{n}}{n!} = e^{J}. $$
Definition (\ref{32}) yields  therefore,
\be
|J ,\gamma,\theta \rangle:=e^{-J/2}\left[e^{-i\theta}\vert \varphi_{00}\rangle +\sum_{n=1}^\infty
\sum_{j=1}^{2} \frac{J^{n/2}e^{-i n \gamma}e^{-ij\theta}}{\sqrt{2n!}}\;\vert\varphi_{nj}\rangle\right]\; .
\label{ex1-cs}
\end{equation}
Actually, this time we can restrict the variable $\gamma$ to the
interval $[0, 2\pi )$ and use the measure $d\mu (\gamma ) =
\displaystyle{\frac 1{2\pi}}\; d\gamma$ in (\ref{gen-meas})
instead of the one in (\ref{ns-scalprod2}). Furthermore, $0 \leq J
< \infty$ and the measure $d\nu (J)$ has to solve the moment
problem
$$ \int_0^\infty J^n\;d\nu (J) = \left\{\begin{matrix} 1\; , \;\;\;\; \text{if}\; n=0\; , \\
                                                       2n!\; ,\;\;\; \text{if} \; n \geq 1 \; .
                                                       \end{matrix}\right. $$
It is then easily seen that $d\nu (J) = [2e^{-J} - \delta (J)]\; dJ .$
Thus, writing
$$dm (J, \gamma , \theta ) = \frac {e^J}{4\pi^2}\;
    [2e^{-J} - \delta (J)]\;dJ\;d\gamma\;d\theta,$$
we can prove the resolution of the identity,
$$
\int \vert J,\gamma ,\theta \rangle\langle J,\gamma ,\theta \vert \;dm (J,\gamma,\theta )= I\; . $$
Finally, introducing the complex variable $z = r e^{-i\gamma} = J^{\frac 12}e^{-i\gamma}, \; z \in
\mathbb C$, we can rewrite (\ref{ex1-cs}) as
\be
|z,\theta \rangle = e^{-\frac {\vert z\vert^2}2}\left[e^{-i\theta}\vert \varphi_{00}\rangle +\sum_{n=1}^\infty
\sum_{j=1}^{2} \frac{z^n \;e^{-ij\theta}}{\sqrt{2n!}}\;\vert\varphi_{nj}\rangle\right]\; .
\label{ex1-cs2}
\end{equation}

\vspace{3mm}

{\bf Example 2:} As a second example consider a particle of mass
$m$ constrained to move on the $xy$-plane and subject to the force
$\vec F=(-kx-by,-ky-bx,0)$, derivable from the potential
$V(x,y)=\frac{1}{2}k(x^2+y^2)+bxy$. In the  rotated coordinates
$\xi_{\pm}=\frac{1}{\sqrt{2}}(x\pm y)$, this potential assumes the
form
$V(\xi_+,\xi_-)=\frac{1}{2}m(\omega_+^2\xi_+^2+\omega_-^2\xi_-^2)$,
where $\omega_\pm^2=\displaystyle{\frac 1m }(k\pm b)$. The
hamiltonian looks like a 2-dimensional harmonic oscillator since,
in an obvious notation, we also have $p_x^2+p_y^2=p_+^2+p_-^2$.
Introducing finally the creation and annihilation operators for
the $\pm$ modes and adding an inessential constant we get
$H=\omega_+ a_+^\dagger a_+ + \omega_- a_-^\dagger a_-$. The
eigenvalues are therefore $E_{n_+,n_-}=\omega_+n_++\omega_-n_-$
and the corresponding eigenstates are
$\varphi_{n_+,n_-}=\displaystyle{\frac{(a_+^\dagger)^{n_+}(a_-^\dagger)^{n_-}}{\sqrt{n_+!n_-!}}}\varphi_{00}$,
where $a_-\varphi_{00}=a_+\varphi_{00}=0$. Let us now take, as a
concrete example, $b=\displaystyle{\frac{3k}{5}}$. Then the
eigenvalues can be written as $E_{n_+,n_-}=\omega_-(2n_++n_-)$ and
the degeneracy can be simply deduced: we notice that the spectrum
is $\omega_- n,\;\; n =2n_++n_-=0,1,2, \ldots\; , $ and
$d(2n)=d(2n+1)=n+1$. Therefore, since $\rho_{2n}=(2n)!(n+1)$ and
$\rho_{2n+1}=(2n+1)!(n+1)$, we may write \be |J,\gamma,\theta
\rangle= e^{-\frac J2}\sum_{l=0}^\infty\sum_{j=1}^{l+1}
\frac{J^le^{-2i\gamma l}e^{-i\theta j}}
{\sqrt{(2l)!(l+1)}}\left[\vert \Psi_{2l,j}\rangle
+\frac{\sqrt{J}e^{-i\gamma}} {\sqrt{2l+1}}\vert
\Psi_{2l+1,j}\rangle \right]\; , \label{ex2-cs1}
\end{equation}
where we have introduced the states $\Psi_{n,j}\;, \;\; n =0,1,2,
\ldots\; ,$ and $j=1,2,\ldots, d(n)$, in order to keep track of
the degeneracy of $H$. It is trivial to check that these states
display temporal stability and the action identity, while it does
not seem to be an easy task to find an explicit expression for a
measure with respect to which a resolution of
the identity would be satisfied. However, as we will discuss in the Appendix, it is
possible to find weight functions, which are not necessarily everywhere
positive, with respect to which a resolution of the identity could be
defined in a {\em weak} sense.

\vspace{3mm}

{\bf Example 3:} Let us consider now a particle of mass $m$ and  electric charge $e$,
subject to a three-dimensional harmonic force $\vec
F=-k(x,y,z)$ and placed in a uniform magnetic field, oriented along the $z$-axis
and given by the vector potential $\vec A=\displaystyle{\frac{B}{2}}(-y,x,0)$. The Hamiltonian
$$ H=\frac{1}{2m}(p_x+\frac{eB}{2}y)^2+
\frac{1}{2m}(p_y-\frac{eB}{2}x)^2+\frac{1}{2m}p_z^2+
\frac{1}{2}k(x^2+y^2+z^2),$$
can be rewritten as
$$H=N_+(\tilde\omega+\Omega)+N_-(\tilde\omega-\Omega)+N_z\omega \; , $$
where we have introduced
\begin{align}
& \qquad\qquad\qquad\Omega=\frac{eB}{2m},\qquad
\omega^2=\frac{k}{m},\qquad
\tilde\omega^2= \Omega^2 + \omega^2,\nonumber\\
& a_u=\frac{1}{\sqrt{2}}\left(\sqrt{m\tilde\omega}u+\frac{i}
{\sqrt{m\tilde\omega}}p_u\right),\;\; u = x, y, \qquad
a_z=\frac{1}{\sqrt{2}}\left(\sqrt{m\omega}z+\frac{i}{\sqrt{m\omega}}p_z\right),\nonumber\\
& \qquad\qquad\qquad  a_\pm=\frac{a_x\pm ia_y}{\sqrt{2}},\qquad  N_\pm=a_\pm^\dagger a_\pm ,\qquad
N_z=a_z^\dagger a_z\; \nonumber.
\end{align}
The eigenvalues and the eigenstates of $H$ are easily found to be
\beano
E_{n_+,n_-,n_z} & = &
n_+(\tilde\omega+\Omega)+n_-(\tilde\omega-\Omega)+n_z\omega\; ,\\
\varphi_{n_+,n_-,n_z} & = &
\frac{(a_+^\dagger)^{n_+}(a_-^\dagger)^{n_-}(a_z^\dagger)^{n_z}}{\sqrt{n_+!n_-!n_z!}}
\varphi_{000}\; ,
\enano
where
$a_-\varphi_{000}=a_+\varphi_{000}=a_z\varphi_{000}=0$. In order
to simplify the computation of the degeneracy of this Hamiltonian
we assume that $\Omega\ll\omega$. In this approximation $H$ can be
written as $H\simeq\omega(N_++N_-+N_z)$, which means that the
eigenvalues really depend only on $n=n_+ +n_- +n_z$. As in the
previous examples we can introduce the eigenvalues $E_n=\omega n$
while the degeneracy of the n-th energy level is
$d(n)=\sum_{k=1}^{n+1}k=\displaystyle{\frac 12}(n+1)(n+2)$. If we denote
the corresponding eigenstates by
$\Psi_{nj}$, $n =0,1,2, \ldots\; , \;\; j=1,2,\ldots\; ,d(n)$,  we find
\be
|J,\gamma, \theta \rangle:= \sqrt{2}\; e^{-\frac J2}\sum_{n=0}^\infty
\sum_{j=1}^{d(n)}\,\frac{J^{n/2}e^{-in
\gamma}e^{-ij\theta}}{\sqrt{(n+2)!}}\;\vert\Psi_{nj}\rangle\; .
\label{ex3-cs}
\end{equation}
Once again, in this case we may introduce the complex variable $z = r e^{-i\gamma} =
J^{\frac 12}e^{-i\gamma}, \; z \in \mathbb C$, and write these coherent states as
\be
|z, \theta \rangle = \sqrt{2}\; e^{-\frac {\vert z \vert^2}2}\sum_{n=0}^\infty
\sum_{j=1}^{d(n)}\,\frac{z^n \;e^{-ij\theta}}{\sqrt{(n+2)!}}\;\vert\Psi_{nj}\rangle\; .
\label{ex3-cs2}
\end{equation}
In this case the resolution of the identity takes the form,
\be
  \frac 1{4\pi^2}\; \int_0^\infty\!\int_0^{2\pi}\!\int_0^{2\pi}\vert z, \theta \rangle
    \langle z, \theta \vert\; r^5\;d\gamma\;d\theta\; dr = I \; .
\label{ex3-resolid}
\end{equation}
It is trivial to check that all the other stated properties are satisfied as well.
We ought to mention here that coherent states for this Hamiltonian have been constructed
before in
\cite{gaznov}. However the treatment there is somewhat different, in that the authors
obtain multidimensional coherent states which allow them to study the
Berezin-Lieb inequalities for the associated thermodynamic potential.

\subsection{Infinite degeracies}\label{inf-deg}

   We are now in a position to construct coherent states for Hamiltonians with infinite degeneracies. Let
$\htil$ be an abstract Hilbert space and $\{\phi_{k\ell}\}_{k,\ell = 0}^\infty$ an orthonormal
basis in it:
$$
  \langle\phi_{k\ell}\mid\phi_{k'\ell'}\rangle = \delta_{kk'}\;\delta_{\ell \ell'}\; . $$
Using these and the basis vectors $\Psi_n$ (see (\ref{ac-ang-bas})) of $\h_{ac-ang}$ we now build
several families of coherent states on $\htil$
\medskip
\begin{itemize}
\item[(1)] Vector coherent states {\bf VCS1}

These are infinite component vector coherent states,
\begin{align}
  \vert J,\gamma;\; J'\gamma'; \; \ell\rangle^1   & =  \displaystyle{\frac {\Psi_\ell (J',\gamma')}
  {[\mathcal N (J) \mathcal N (J')]^{\frac 12}}}\sum_{n=0}^\infty\overline{\Psi_n (J, \gamma )}
  \vert\phi_{n\ell}\rangle\nonumber\\
  & =  \displaystyle{\frac {J^{\prime\frac {\ell}2}\;e^{i\epsilon_\ell \gamma'} }
  {[\mathcal N (J) \mathcal N (J')]^{\frac 12}}}\sum_{n=0}^\infty\frac {J^{\frac n2}\;e^{-i\epsilon_n \gamma}}
  {[\epsilon_\ell ! \epsilon_n !]^{\frac 12}}\vert \phi_{n\ell}\rangle \; ,
\label{VCS1}
\end{align}
with components $\ell =0,1,2, \ldots \; .$ These vectors satisfy the normalization
$$ \sum_{\ell = 0}^{\infty}  \; ^1\!\langle J, \gamma ; \; J', \gamma';\; \ell \mid
    J, \gamma ; \; J', \gamma';\; \ell \rangle^1 = 1\; ,  $$
(note that according to our present convention, the individual vectors are not
normalized) and the resolution of identity condition,
\begin{align}
 & \sum_{\ell = 0}^\infty\int_{\mathbb R^+}\!\int_{\mathbb R^+}\!\bigg[\int_0^L\!\int_0^L\!
   \vert J, \gamma ; \; J', \gamma';\; \ell \rangle^1 \;^1 \!\langle J, \gamma ; \; J', \gamma';\; \ell\vert
   \nonumber\\
   & \qquad\qquad \times d\mu(\gamma )\; d\mu(\gamma' )\bigg]\; \mathcal N (J) \mathcal N (J')
   \; d\nu (J)\; d\nu (J')
   = I_{\htil} \; .
\label{VCS11}
\end{align}
Consider now the Hamiltonian
\be
  H_1 = \sum_{n,\ell =0}^\infty \omega\epsilon_n \vert\phi_{n\ell}\rangle\langle \phi_{n\ell}\vert
  = \omega A_1^\dagger A_1 \; ,
\label{VCS-hamilt1}
\end{equation}
where $A_1 , A_1^\dagger$ are the operators
\be
  A_1\phi_{n\ell} = \sqrt{\epsilon_n}\;\phi_{n-1\; \ell}  \; , \qquad
  A_1^\dagger\phi_{n\ell} = \sqrt{\epsilon_{n+1}}\;\phi_{n+1\; \ell} \; .
\label{A-Adag1}
\end{equation}
Each level $\omega\epsilon_n$ of this Hamiltonian is infinitely degenerate, with $\ell$ counting the
degeneracy. Thus the states (\ref{VCS1}) are  Gazeau-Klauder type vector coherent states for this Hamiltonian.
Indeed, they satisfy the time stability condition,
\be
  e^{-iH_1 t}\vert J,\gamma;\; J'\gamma'; \; \ell\rangle^1  =
    \vert J,\gamma + \omega t ;\; J'\gamma'; \; \ell\rangle^1\; ,
\label{time-stab1}
\end{equation}
and an action identity, which we could write either as
\be
  \frac{\;^1\!\langle J,\gamma;\; J'\gamma'; \; \ell\mid H_1\mid J,\gamma;\; J'\gamma'; \; \ell\rangle^1}
   {\Vert \vert J,\gamma;\; J'\gamma'; \; \ell\rangle^1\Vert^2} = \omega J\; ,
\label{act-id1}
\end{equation}
or as
\be
\sum_{\ell = 0}^{\infty}  \; ^1\!\langle J, \gamma ; \; J', \gamma';\; \ell \mid H_1 \mid
    J, \gamma ; \; J', \gamma';\; \ell \rangle^1 = \omega J\; ,
\label{act-id2}
\end{equation}
where we have summed over the degenerate levels.

   Note that we could just as well have constructed vector coherent states in this example, using an
orthonormal basis $\{\Psi_n\}_{n=0}^\infty$ in an arbitrary reproducing kernel Hilbert space
$\h_\text{ker}$:
\be
 \vert J, \gamma ; \; x;\; \ell \rangle =  \displaystyle{\frac {\Psi_\ell (x)}
  {[K(x,x)\mathcal N (J')]^{\frac 12}}}\sum_{n=0}^\infty
  \frac {J^{\frac n2}\;e^{-i\epsilon_n \gamma}}
  {[\epsilon_\ell ! \epsilon_n !]^{\frac 12}}\vert \phi_{n\ell}\rangle \; ,
\label{alt-vcs}
\end{equation}
with $K(x,x)$ as in (\ref{repkercond}) and the degeneracies would again be handled as before. However,
the special choice made in (\ref{VCS1}) enables us to write down the related family of vector coherent
states, appearing in (\ref{VCS2}) below, which are the coherent states of a second Hamiltonian, acting
on the degeneracy levels.

\medskip

\item[(2)] Vector coherent states {\bf VCS2}

These are a second set of similar vector coherent states
\begin{align}
  \vert J,\gamma;\; J'\gamma';\; n\rangle^2  &=  \displaystyle{\frac {\overline{\Psi_n (J,\gamma)}}
  {[\mathcal N (J) \mathcal N (J')]^{\frac 12}}}\sum_{\ell =0}^\infty\Psi_\ell (J', \gamma' )
  \vert\phi_{n\ell}\rangle\nonumber\\
  &=  \displaystyle{\frac {J^{\frac n2}\;e^{-i\epsilon_n \gamma} }
  {[\mathcal N (J) \mathcal N (J')]^{\frac 12}}}\sum_{\ell=0}^\infty
  \frac {J^{\prime\frac \ell 2}\;e^{i\epsilon_\ell \gamma'}}
  {[\epsilon_\ell ! \epsilon_n !]^{\frac 12}}\vert\phi_{n\ell}\rangle\; ,
\label{VCS2}
\end{align}
with components $n =0,1,2, \ldots \; .$ Defining a second Hamiltonian,
\be
  H_2 = \sum_{n,\ell =0}^\infty \omega\epsilon_\ell \vert\phi_{n\ell}\rangle\langle \phi_{n\ell}\vert
  = \omega A_2^\dagger A_2 \; ,
\label{VCS-hamilt2}
\end{equation}
where $A_2 , A_2^\dagger$ are the operators
\be
  A_2\phi_{n\ell} = \sqrt{\epsilon_\ell}\;\phi_{n\; \ell - 1}  \; , \qquad
  A_2^\dagger\phi_{n\ell} = \sqrt{\epsilon_{\ell+1}}\;\phi_{n\; \ell + 1} \; ,
\label{A-Adag1bis}
\end{equation}
we see that the states (\ref{VCS2}) are Kazeau-Klauder type  coherent states states for this Hamiltonian.
The two Hamiltonians $H_1$ and $H_2$ commute and, in fact, $H_2$ lifts the degeracy of $H_1$ and vice versa.

Finally, we can define a third set of coherent states as below.

\medskip

\item[(3)] `Bi-coherent states' {\bf BCS}

These are basically the summed-over versions of the previous two
\begin{align}
   \vert J,\gamma;\; J',\gamma'\rangle^\text{BCS}   & =  \displaystyle{\frac 1
  {[\mathcal N (J) \mathcal N (J')]^{\frac 12}}}\sum_{n,\ell =0}^\infty
   \overline{\Psi_n (J,\gamma)}\Psi_\ell (J', \gamma' )
  \vert\phi_{n\ell}\rangle\nonumber\\
   & =  \displaystyle{\frac 1
  {[\mathcal N (J) \mathcal N (J')]^{\frac 12}}}\sum_{n,\ell=0}^\infty
  \frac {J^{\frac n2}J^{\prime\frac \ell 2}\;e^{-i(\epsilon_n \gamma - \epsilon_\ell \gamma' )}}
  {[\epsilon_n ! \epsilon_\ell !]^{\frac 12}}\vert\phi_{n\ell}\rangle\; ,
\label{BVCS}
\end{align}
which can be considered as being the multidimensional  coherent states
(see \cite{gaznov}) of the Hamiltonian
\be
  H = H_1 -H_2  = \sum_{n,\ell =0}^\infty \omega(\epsilon_n - \epsilon_\ell )
     \vert\phi_{n\ell}\rangle\langle \phi_{n\ell}\vert
  = \omega[ A_1^\dagger A_1 -  A_2^\dagger A_2 ]\; .
\label{bi-hamilt}
\end{equation}
These coherent states are normalized to unity; they satisfy the resolution of the identity,
 \begin{align}
 & \int_{\mathbb R^+}\!\int_{\mathbb R^+}\!\bigg[\int_0^L\!\int_0^L\;
   \vert J, \gamma ; \; J', \gamma'\rangle\!^\text{BCS}\;^\text{BCS}\!\langle J, \gamma ; \; J', \gamma'\vert
   \nonumber\\
   &\qquad\qquad  \times d\mu(\gamma )\; d\mu(\gamma' )\bigg]
   \mathcal N (J)\mathcal N (J')\;d\nu (J)\; d\nu (J') = I_{\htil} \; ,
\label{bi-hamilt2}
\end{align}
temporal stability condition,
\begin{equation}
 e^{-iHt}\vert J,\gamma;\; J',\gamma'\rangle^\text{BCS} =
  \vert J,\gamma + \omega t ;\; J',\gamma' + \omega t\rangle^\text{BCS} \;,
\label{bi-hamilt3}
\end{equation}
and the action identity,
\be
  \;^\text{BCS}\!\langle J, \gamma ; \; J', \gamma'\vert H\vert J, \gamma ; \; J', \gamma'\rangle\!^\text{BCS}
   = \omega (J - J')\; .
\label{BCS-act-id}
\end{equation}

\end{itemize}

A physical example of a Hamiltonian admitting such infinite degeneracies is worked out in the following
section.

\section{Electron in a magnetic field}\label{elec-mag-fld}
  A single electron of unit charge, placed in the $xy$-plane and subjected to a constant magnetic
field, pointing along the negative $z$-direction, has the classical Hamiltonian
\be
  H_\text{elec} = \frac 12 (\vec p + \vec A )^2 = \frac 12 \left(p_x + \frac y2 \right)^2 +
      \frac 12 \left(p_y - \frac x2 \right)^2\; ,
\label{elec-ham}
\end{equation}
where we have chosen the magnetic vector potential to be $\vec A = \frac 12 (y, -x, 0)$,
using the convenient units introduced in \cite{BMS}. On
$\htil = L^2 (\mathbb R^2 , dxdy )$ we introduce the quantized observables,
\be
 p_x + \frac y2 \longrightarrow Q_1 = -i\frac \partial{\partial x } + \frac y2 \; , \qquad
 p_y - \frac x2 \longrightarrow P_1 = -i\frac \partial{\partial y } - \frac x2 \; ,
\label{quant-obs1}
\end{equation}
which satisfy $[Q_1, P_1 ] = iI_{\htil}$ and in terms of which the quantum Hamiltonian,
corresponding to $H_\text{elec}$ becomes
\be
  H_1 = \frac 12 \left( P_1^2 + Q_1^2\right)\; .
\label{qu-elec-ham1}
\end{equation}
This is just the oscillator Hamiltonian in one dimension, with eigenvalues $E_n = \omega (n + \frac 12 ), \;
n =0,1,2, \ldots \infty$.
Each level is infinitely degenerate, and we will denote the corresponding normalized eigenvectors by
$\Psi_{n \ell} ,\;\; \ell = 0,1,2, \ldots , \infty$. If the magnetic field were aligned along the positive
$z$-axis (with $\vec A = \frac 12 (-y, x, 0)$), the corresponding quantum Hamiltonian would have been
\be
  H_2 = \frac 12 \left( P_2^2 + Q_2^2\right)\; .
\label{qu-elec-ham2}
\end{equation}
with
\be
 Q_2 = -i\frac \partial{\partial y } + \frac x2 \; , \qquad
 P_2 = -i\frac \partial{\partial x } - \frac y2 \; ,
\label{quant-obs2}
\end{equation}
and $[Q_2, P_2 ] = iI_{\htil}$. The two sets of operators $\{ Q_i , P_i\}, \; i =1,2$, mutually commute:
\be
  [Q_1 , Q_2 ] = [Q_1 , P_2 ] = [P_1 , Q_2 ] = [P_1 , P_2 ] = 0 \; .
\label{commutants}
\end{equation}
(Note that at the classical level, the transformation $(x, y , p_x , p_y ) \longrightarrow (x' = p_x + \frac y2 ,
\; y' = p_y + \frac x2 , \; p_{x'} = p_y - \frac x2 , \; p_{y'} =  p_y - \frac y2 $) is  canonical, i.e.,
$dx\wedge dp_x + dy\wedge dp_y = dx'\wedge dp_{x'} + dy'\wedge dp_{y'}$).
Thus, $[H_1 , H_2 ] = 0$ and the eigenvectors $\Psi_{n \ell}$ of $H_1$ can be so chosen that they are
also the eigenvectors of $H_2$ in the manner
\be
  H_1\Psi_{n\ell} = \omega (n + \frac 12 )\Psi_{n\ell} \; , \qquad
  H_2\Psi_{n\ell} = \omega (\ell + \frac 12 )\Psi_{n\ell}\; ,
\label{deg-lift}
\end{equation}
so that $H_2$ lifts the degeneracy of $H_1$ and vice versa. We shall assume that this has been done.

   While we shall follow the technique outlined in the previous section to construct vector coherent
states for the above two Hamiltonians, we shall first analyze the algebraic structures generated by these
operators, to get a deeper insight into the nature of the resulting coherent states. In the process
we shall display some von Neumann algebraic properties, the appearance of KMS states  and a certain {\em modular
structure} carried by the above model. Details of the mathematical theory underlying these structures
may be found in  \cite{AAG-book,aliemch,hahuwi,tak1,tak2}.   On
$\h = L^2 (\mathbb R )$ let $Q$ and $P$ be the usual position and momentum operators in the
Schr\"odinger representation. Denote by $\mathcal B_2 (\h) \simeq \h\otimes \overline{\h}$ the space of
Hilbert-Schmidt operators on $\h$. This is again a Hilbert space, with the scalar product
$\langle X\mid Y\rangle_2 = \text{Tr} [X^* Y]$. Let $\{\phi_n\}_{n=0}^\infty$ be the orthonormal
basis of $\h$ consisting of the eigenvectors of the oscillator Hamiltonian $H_\text{osc} =
\frac 12 (P^2 + Q^2 )$, i.e., $H_\text{osc}\phi_n = \omega (n + \frac 12 )\phi_n , \; n =0,1,2, \ldots$. Then,
\be
    \phi_{n\ell} := \vert\phi_n\rangle \langle \phi_\ell \vert , \qquad n,\ell = 0,1,2, \ldots , \infty,
\label{hilb-sch-bas}
\end{equation}
is an orthonormal basis for $\mathcal B_2 (\h)$. On $\h$ define the unitary operators,
\be
 U(x, y) = e^{-i(xQ + yP)}, \; \quad (U(x,y)\phi )(\xi) = e^{-ix (\xi -\frac y2 )}\phi (\xi -y ), \quad
    (x,y) \in \mathbb R^2 , \;\; \phi \in \h \; .
\label{CCR-rep}
\end{equation}
Then, it is well known (see, for example, \cite{AAG-book}) that the  map,
\be
 \mathcal W : \mathcal B_2 (\h ) \longrightarrow L^2 (\mathbb R^2 , dx\; dy ) = \htil , \qquad
(\mathcal W X)(x,y) = \frac 1{(2\pi)^\frac 12}\text{Tr} [U(x,y)^* X]\; ,
\label{wigmap}
\end{equation}
is unitary. Next, if $A$ and $B$ are two operators on $\h$, we define by $A\vee B$ the operator
$$ A\vee B (X) = AXB^* , \qquad X \in \mathcal B_2 (\h )\; . $$
For a large class of operators $A, B$ (in particular when $A$ and  $B$ are both bounded operators),
$A\vee B$ defines a linear operator on $\mathcal B_2 (\h )$. Then straightforward computations
(as shown in the Appendix) yield,
\be
 \mathcal W \begin{pmatrix} Q\vee I_\h \\ P\vee I_\h \end{pmatrix}\mathcal W ^{-1} =
 \begin{pmatrix} Q_1 \\ P_1 \end{pmatrix}\; , \qquad
 \mathcal W \begin{pmatrix} I_\h \vee Q \\ I_\h \vee P \end{pmatrix} \mathcal W^{-1} =
 \begin{pmatrix} P_2 \\ Q_2 \end{pmatrix}\;,
\label{alg-transf1}
\end{equation}
and
\be
\mathcal W \begin{pmatrix} H_\text{osc}\vee I_\h \\ I_\h\vee H_\text{osc} \end{pmatrix}\mathcal W^{-1} =
 \begin{pmatrix} H_1 \\ H_2 \end{pmatrix}\; , \qquad \mathcal W \phi_{n\ell} = \Psi_{n\ell },
\label{alg-transf2}
\end{equation}
where the $\phi_{n\ell}$ are the basis vectors defined in (\ref{hilb-sch-bas}) and the $\Psi_{n\ell}$ are the
normalized eigenvectors defined in (\ref{deg-lift}). This also means that these latter vectors form a
basis of $L^2(\mathbb R^2 , dxdy )$.

  In the sequel we shall also need the thermal equilibrium state, at inverse temperature $\beta$,
corresponding to the Hamiltonian $H_\text{osc}$. This is the density matrix,
\be
  \rho_\beta = \frac {e^{-\beta H_\text{osc}}}{\text{Tr}\left[e^{-\beta H_\text{osc}}\right]} =
     (1- e^{-\omega\beta})\sum_{n=0}^\infty e^{-n\omega\beta}
     \vert\phi_n \rangle\langle\phi_n\vert\; .
\label{osc-therm-st}
\end{equation}

   On $\htil$, for each $(x,y) \in \mathbb R^2$, define the operators
\be
  U_1 (x,y) = \mathcal W \;[U(x,y)\vee I_\h ]\;\mathcal W^{-1} \; , \quad
  U_2 (x,y) = \mathcal W \;[I_\h \vee U(x,y)^*]\; \mathcal W^{-1} \;,
\label{bi-alg1}
\end{equation}
and let $\mathfrak A_i\; i=1,2$, be the von-Neumann algebra (see, e.g. \cite{tak1})
generated by the unitary operators
$\{U_i (x,y)\mid (x,y) \in \mathbb R^2\} .$ Then  using the unitary map $\mathcal W$,
the following modular structure
can easily be inferred for the pair of von Neumann algebras $\mathfrak A_1$ and
$\mathfrak A_2$ (for details on modular structures see \cite{tak2} and for the particular
type of algebras appearing here, see
\cite{AAG-book,aliemch}).

\bei
\item[(1)] The algebra $\mathfrak A_1$ is the commutant of the algebra $\mathfrak A_2$ and vice versa
and $\mathfrak A_1 \cap \;\mathfrak A_2 = \mathbb C I_{\htil}$.

\medskip

\item[(2)] If $\{\lambda_n\}_{n=0}^\infty$ is a sequence of non-zero positive numbers such that
$\sum_{n=0}^\infty \lambda_n =1$, then the vector $\Phi = \sum_{n=0}^\infty \lambda_n^{\frac 12}
\Psi_{nn}$ is cyclic and separating for $\mathfrak A_1$. In particular, we shall work with the
vector $\Phi = \Phi_\beta$, for which the  $\lambda_n$ correspond to the thermal state $\rho_\beta$ in
(\ref{osc-therm-st}):
\be
  \Phi_\beta =  [1- e^{-\omega\beta}]^{\frac 12}\sum_{n=0}^\infty e^{-\frac{n\omega\beta}2}
     \Psi_{nn} \; , \quad \text{i.e.,} \quad  \lambda_n = (1- e^{-\omega\beta})e^{-n\omega\beta}\; .
\label{KMS-st1}
\end{equation}

\medskip

\item[(3)] The map
\be
  S_\beta : \htil \longrightarrow \htil\; , \qquad
  S_\beta [U_1 (x,y)\Phi_\beta] = U_1 (x,y)^*\Phi_\beta\; ,
\label{antunit-map1}
\end{equation}
is closable and has the polar decomposition,
\be
  S_\beta = J_\beta \; \Delta_\beta^\frac 12\;,
\label{polar-dec}
\end{equation}
where $J_\beta$ is the {\em antiunitary} operator:
\be
  J_\beta\Psi_{n\ell} = \Psi_{\ell n}\; ,\qquad  J_\beta^2 = I_{\htil}\; , \qquad J_\beta \Phi_\beta
   = \Phi_\beta\; ,
\label{antiunit-map2}
\end{equation}
so that $J_\beta \mathfrak A_1 J_\beta = \mathfrak A_2$, and $\Delta_\beta$ is the self-adjoint operator,
\be
  \Delta_\beta = \sum_{n,\ell = 0}^\infty \frac {\lambda_n}{\lambda_\ell}\vert\Psi_{n\ell}\rangle
    \langle\Psi_{n\ell}\vert = e^{-\beta H}\; \quad \text{where} \quad H = H_1 - H_2\; ,
\label{antiunit-map3}
\end{equation}
the Hamiltonians $H_1$ and $H_2$ being as in (\ref{deg-lift}). (We reproduce the derivation of
(\ref{polar-dec})-(\ref{antiunit-map3}) in the Appendix).  The  operator $\Delta_\beta$ defines a one parameter
group of evolution, $t \longmapsto \alpha_\beta (t)$ on the algebra $\mathfrak A_1$:
\be
\alpha_\beta (t)[A] = \Delta_\beta^{-\frac {it}\beta} A \;\Delta_\beta^{\frac {it}\beta} =
            e^{itH}A\; e^{-itH} = e^{itH_1}A\; e^{-itH_1}\; , \qquad A \in \mathfrak A_1\; .
\label{KMS-evol}
\end{equation}

\medskip

\item[(4)] The state $\varphi_\beta$, defined on the algebra $\mathfrak A_1$ by the vector $\Phi_\beta$:
\be
   \langle\varphi_\beta\; ; \; A\rangle = \langle \Phi_\beta \mid A\Phi_\beta \rangle_{\htil}\; ,
   \qquad A\in \mathfrak A_1\; ,
\label{KMS-vect-st}
\end{equation}
is a faithful normal vector state which is invariant under the evolution $\alpha_\beta$:
\be
  \langle\varphi_\beta\; ; \; \alpha_\beta (t) [A]\rangle = \langle\varphi_\beta\; ; \; A\rangle\; .
\label{KMS-inv}
\end{equation}
Furthermore, $\varphi_\beta$ is a KMS state \cite{hahuwi,tak2} in the following sense:
for $A, B \in \mathfrak A_1$, define the
function $F_{A,B}$ of the real variable $t$,
\be
F_{A,B} (t) = \langle \varphi_\beta\; ;\; A\alpha_\beta (t)[B]\rangle\; .
\label{KMS-cond1}
\end{equation}
Then this function has an analytic extension to the open strip $\{ z = t +iv \mid 0 < v < \beta \}$ and
furthermore,
\be
   F_{A,B} (t + i\beta ) = \langle\varphi_\beta\; ;\; \alpha_\beta (t) [B]A\rangle\;  .
\label{KMS-cond2}
\end{equation}
\eni

\medskip

Going back now to the problem of constructing coherent states for this system, we can immediately
write down three types of states, in analogy with (\ref{VCS1}), (\ref{VCS2}) and (\ref{BVCS}).

\bei
\item[(1)] Vector coherent states of the Hamiltonian $H_1 - \frac {\omega}2 I_{\htil}$

These are the states on $\htil = L^2 (\mathbb R^2 , dx\; dy )$,
\be
  \vert z, \overline{z}';\; \ell\rangle^1 = e^{-\frac {\vert z\vert^2 + \vert z'\vert^2}2}
  \;\overline{z}^{\prime \ell}
  \;\sum_{n=0}^\infty \frac {z^n}{\sqrt{n! \;\ell !}}\;\vert\Psi_{n\ell}\rangle\; , \qquad \ell = 0, 1,2,
  \ldots , \infty\; .
\label{vcs-h1}
\end{equation}
They are obtained by replacing  $J^{\frac n2}e^{-i\epsilon_n \gamma}$ by $z^n = r^n e^{i\ell\theta}$ and
$J^{\prime\frac \ell2} e^{-i\epsilon_\ell\gamma'}$ by $z^{\prime\ell} = r^{\prime\ell} e^{i\ell
\theta'}$ in
(\ref{VCS1}), with $z,z' \in \mathbb C$. The resolution of the
identity now takes the form:
\be
\frac 1{(2\pi )^2}
  \sum_{\ell =0}^\infty\int_{\mathbb C \times \mathbb C}\vert z, \overline{z}';\; \ell\rangle\!^1\;^1\!\langle
    z,\overline{z}';\; \ell \vert\; dx\;dx'\;dy\;dy' = I_{\htil}\; , \qquad
\label{KMS-resolid1}
\end{equation}
where $z = \frac 1{\sqrt{2}}(y - ix)$ and $z' = \frac 1{\sqrt{2}}( y' - ix')\;$.
Let us introduce
the operators,
\be
  A_\ell = \frac 1{\sqrt{2}} (Q_\ell + iP_\ell) \; , \quad A_\ell^\dag  =
  \frac 1{\sqrt{2}} (Q_\ell - iP_\ell)\; ,
  \quad H_\ell = A^\dag_\ell A_\ell + \frac {\omega}2\; , \quad \ell =1,2 \; .
\label{CCR-ops1}
\end{equation}
Then, it is not hard to see that,
\be
 U_1 (z ) := U_1(x,y) = e^{zA_1^\dag -\overline{z}A_1} = e^{-\frac 12 \vert z\vert^2}
   e^{zA_1^\dag}e^{ -\overline{z}A_1}\;.
\label{CCR-ops2}
\end{equation}
Also, since
$$ A_1 \vert \Psi_{n \ell}\rangle = \sqrt{n}\vert \Psi_{n-1\; \ell}\rangle\; , \qquad
   A_1^\dag \vert \Psi_{n \ell}\rangle = \sqrt{n+1}\vert \Psi_{n+1 \; \ell}\rangle\; , $$
it  easily follows that,
\be
  \vert z, \overline{z}';\; \ell\rangle^1 = e^{-\frac {\vert z'\vert^2}2}
  \;\frac {\overline{z}^{\prime \ell}}{\sqrt {\ell !}}\; U_1 (z)\vert \Psi_{0 \ell}\rangle\; .
\label{VCS11bis}
\end{equation}

\medskip

\item[(2)] Vector coherent states of the Hamiltonian $H_2 - \frac {\omega}2 I_{\htil}$

Following (\ref{VCS2}), we have the analogous set of vector coherent states
\bea
  \vert z, \overline{z}';\; n\rangle^2 & = &  e^{-\frac {\vert z\vert^2 + \vert z'\vert^2}2}
  \;z^n
  \;\sum_{\ell =0}^\infty \frac {\overline{z}^{\prime\ell}}{\sqrt{n! \;\ell !}}\;\vert\Psi_{n\ell}\rangle
  \nonumber\\
   & = & e^{-\frac {\vert z\vert^2}2}
  \;\frac {z^n}{\sqrt {n !}}\; U_2 (z')\vert \Psi_{n 0}\rangle \; ,
  \qquad n = 0,1,2, \ldots , \infty\; .
\label{vcs-h2}
\ena
which satisfy a resolution of the identity similar to (\ref{KMS-resolid1}).

\medskip

\item[(3)] Coherent states of the Hamiltonian $H = H_1 - H_2$

These are the ``bi-coherent states'', analogous to (\ref{BVCS}),
\be
\vert z, \overline{z}'\rangle^\text{BCS} = e^{-\frac {\vert z\vert^2 + \vert z'\vert^2}2}
  \;\sum_{n, \ell =0}^\infty \frac {z^n \overline{z}^{\prime \ell}}{\sqrt{n! \;\ell !}}\;
\vert\Psi_{n\ell}\rangle = U_1 (z) U_2 (z')\vert\Psi_{00}\rangle\; .
\label{BVCS-h}
\end{equation}

\medskip

\item[(4)] Coherent states built from the thermal equilibrium state.

As yet another example related to this system, we build coherent states, starting with
the thermal state $\Phi_\beta$ (see (\ref{KMS-st1}) and  (\ref{KMS-vect-st})). We define these
states as
\be
  \vert z, \overline{z}, \beta\rangle\!^\text{KMS} = U_1 (z) \vert \Phi_\beta \rangle =
              e^{zA_1^\dag - \overline{z}A_1} \vert\Phi_\beta \rangle\; .
\label{KMS-CS}
\end{equation}
In view of the fact that for any normalized vector $\phi \in \h$, the vectors
$U(z)\phi , \;\; z \in \mathbb C$, where $U(z) := U(x,y)$ (see (\ref{CCR-rep})),
satisfy
$$
 \frac 1{2\pi}\;\int_{\mathbb C} \vert U(z)\phi\rangle\langle U(z)\phi\vert\; dx\; dy = I_\h , $$
we deduce, using the isometry $\mathcal W$ in (\ref{wigmap}) that the coherent states
(\ref{KMS-CS}) satisfy the resolution of the identity
condition
\be
  \frac 1{2\pi}\;\int_{\mathbb C}\vert z, \overline{z}, \beta\rangle\!^\text{KMS}\;^\text{KMS}\!
    \langle z, \overline{z}, \beta \vert\; dx\; dy = I_{\htil}\; .
\label{KMS-CS-resolid}
\end{equation}
Also, since
\be
  U_1 (z)\vert\Psi_{nn}\rangle = \frac 1{\sqrt{n!}}\;(A_1^\dag - \overline{z}I_{\htil} )^n \;U_1 (z)
  \vert \Psi_{0n}\rangle = \frac 1{\sqrt{n!}}\;\left(\frac \partial{\partial z } -
  \frac {\overline{z}}2 \;I_{\htil} \right)^n U_1 (z)\vert \Psi_{0n}\rangle\; ,
\label{KMS-CS2}
\end{equation}
which follows from the fact that
$$(A_1^\dag )^n\vert\Psi_{0n}\rangle = \sqrt{n!}\vert\Psi_{nn}\rangle\; ,
\qquad \text{and} \qquad  U_1 (z)\vert\Psi_{0n}\rangle = e^{-\frac {\vert z\vert^2}2}\; \sum_{k =0}^\infty
   \frac {(zA_1)^k}{k!}\vert \Psi_{0n}\rangle, $$
we may rewrite (\ref{KMS-CS}) as
\be
\vert z, \overline{z}, \beta\rangle\!^\text{KMS} = \left[1 - e^{-\omega\beta}\right]^{\frac 12}\;
\sum_{n=0}^\infty \sqrt{n!}\;e^{-\frac {n\omega\beta}2}\;\left(\frac \partial{\partial z}
- \frac {\overline{z}}2\right)^n
  \vert z; \; n\rangle \; ,
\label{KMS-CS3}
\end{equation}
where we have set
 $$ \vert z;\; n\rangle = U_1(z)\vert \Psi_{0n}\rangle\; . $$
Furthermore, using the fact that
$$ (A_2^\dag )^n\vert\Psi_{n0}\rangle = \sqrt{n!}\vert\Psi_{nn}\rangle\; , $$
we may also write
\be
\vert z, \overline{z}, \beta\rangle\!^\text{KMS} = \left[1 - e^{-\omega\beta}\right]^{\frac 12}\;
\sum_{n=0}^\infty \;e^{-\frac {n\omega\beta}2}\;\left(\frac \partial{\partial z}
- \frac {\overline{z}}2\right)^n A_2^n
  \vert z; \; 0\rangle \; ,
\label{KMS-CS4}
\end{equation}
It ought to be pointed out that the coherent states (\ref{KMS-CS}) are not of the Gazeau-Klauder type.
States of the type
$$\left(\frac \partial{\partial z} - \frac {\overline{z}}2\right)^n \vert z; \; n\rangle
= (A_1^\dag - \overline{z}I_{\htil} )^n\vert z; \; n\rangle\; , $$
are finite linear combinations of {\em photon-added coherent states} (see \cite{agtar}),
which have been studied extensively in the optical literature. Note that
\be
  \langle z;\; n\mid z';\; m\rangle = e^{-\frac {\vert z\vert^2 + \vert z'\vert^2}2}\;
     e^{\overline{z}z'}\; \delta_{nm}\; .
\label{bin-orth}
\end{equation}

  Finally, note that since $U_1 (x, y )^* = U_1 (-x, -y )$, using (\ref{antunit-map1})
we can get another family of
coherent states built on the thermal state $\Phi_\beta$:
$$ S_\beta \vert z, \overline{z} , \beta\rangle^\text{KMS} =
  \vert -z, -\overline{z} , \beta\rangle^\text{KMS}\; .$$
Obviously, these also satisfy the same resolution of the identity as (\ref{KMS-CS-resolid}).
\eni

We shall consider in more detail the relationship between the
above algebraic structure and the different kinds of coherent states
discussed here, as well as their  use in the analysis of the
quantum Hall effect, in a subsequent paper.

\section{Appendix}\label{sec-app}
We work out here some of the results quoted in the last  two sections.

\subsection{The measure in Example 2 of Section \ref{fin-deg}}

The proof of the existence of the measure in Example 2 of Section \ref{fin-deg}
will be considered as a particular case of a more general situation.

We are looking for a ``density'' $f(x)$ such that, given a sequence of
numbers $\rho_n$, the following equation holds:
$$
\int_0^\infty f(x)x^ndx=\rho_n,\hspace{3cm} n=0,1,2,\ldots
$$
It is convenient to introduce a new function $\tilde f(x)$ as
$f(x)=e^{-x}\tilde f(x)$ and restate the problem as follows: we are looking
for a function $\tilde f(x)$ such that \be\label{1}
\int_0^\infty \tilde f(x)x^n(e^{-x}dx)=\rho_n,\hspace{3cm}
n=0,1,2,\ldots\; .
 \en
As  is well known, the orthonormalization
procedure in $L^2(\mathbb R^+, e^{-x}dx)$ for $x^n$ produces
the Laguerre polynomials:
\be\label{2}
x^n \rightarrow
L_n(x)=\sum_{k=0}^n \left(
\begin{array}{c}
  n \\
  n-k \\
\end{array}
\right)\frac{(-1)^k}{k!}x^k,
\en
and $<L_n \mid L_l>=\delta_{nl}$,
where the scalar product is, of course, the one in
$L^2(\mathbb R^+, e^{-x}dx)$. If we  consider the linear combination of
(\ref{1}) with the coefficients given in (\ref{2}) we get
\be\label{3}
\int_0^\infty \tilde f(x)L_n(x)(e^{-x}dx)=
\sum_{k=0}^n\left(%
\begin{array}{c}
  n \\
  n-k \\
\end{array}%
\right)\frac{(-1)^k}{k!}\rho_k=:d_n
\en
It is clear then that we
have to take $\tilde f(x)=\sum_{n=0}^\infty\,d_nL_n(x)$, provided this sum
converges and consequently, the required ``density'' is
$f(x)=e^{-x}\left(\sum_{n=0}^\infty\,d_nL_n(x)\right)$. Note however, that this
function is not everywhere positive.

  We can say more on the
coefficients $d_n$ by recalling that $\rho_{2n}=(2n)!(n+1)$
and $\rho_{2n+1}=(2n+1)!(n+1)$. It is an easy exercise to check
that $$
d_n=\sum_{l=0}^{[n/2]}\left(%
\begin{array}{c}
  n \\
  n-2l \\
\end{array}%
\right)(l+1)-\sum_{l=0}^{[(n-1)/2]}\left(%
\begin{array}{c}
  n \\
  n-(2l+1) \\
\end{array}%
\right)(l+1), $$ where  $[r]$ stands for the integer part of the
rational number $r$.  This implies that $d_1=0$ and $d_n=2^{n-2}$
for all $n\geq 2$, so that $\tilde f(x)$ cannot be a
square-integrable function. However, if we consider the sequence
$\{\tilde f_N \;\vert\;  N\in\mathbb N\}$, where  $\tilde
f_N(x)=\sum_{n=0}^N\,d_nL_n(x)$, it is possible to show that it
converges with respect to  a certain family of test functions. For that we
define
\be
{\mathcal D}_b=\left\{f \in{\cal D}([0,1]) \;\; \left|\;\;\;\;
 \int_0^1 \left| \frac {d^k}{dx^k}f(x)\right|\,dx\leq 1,\right.\,\,\forall
k=0,1,2,\ldots\right\}.
\en
This is a non empty subset of ${\mathcal D}([0,1])$. We can check that
\be
I_{NM}:=\int_0^\infty(\tilde
f_N(x)-\tilde f_M(x))\varphi(x)dx\rightarrow 0,
\en
as $N,M\rightarrow \infty$ for all $\varphi\in{\cal D}_b$. This
follows from the fact that
$$L_n(x)=\displaystyle{\frac{1}{n!}\; e^x\frac{d^n}{dx^n}(e^{-x}x^n)}$$
and from the properties of ${\cal D}_b$. Thus,  using integration by parts:
$$
|I_{NM}|\leq
\sum_{n=M+1}^N\, \frac {|d_n|}{n!}\,\int_0^1\left|x^n\left(1+\frac{d}{dx}\right)^n\varphi(x)\right|\,dx\leq
\sum_{n=M+1}^N\,\frac{2^{n-2}\cdot2^n}{n!}\rightarrow0,
$$
as $N,M\rightarrow \infty$.

It may be worth remarking that the set ${\mathcal D}_b$ could be
replaced by some larger set without affecting the final result. However,
the estimates above would have been harder to obtain.  Thus,
since such a  stronger result would not be very relevant in the present
context, we will not consider this generalization here.

\subsection{Proof of (\ref{alg-transf1})}
We only demonstrate the first two relations in (\ref{alg-transf1}), since the other two follow in an entirely
analogous manner. Moreover, (\ref{alg-transf2}) is a direct consequence of (\ref{alg-transf1}).
Consider $X \in \mathcal B_2 (\h )$ of the type $X = \vert\phi\rangle\langle\psi\vert$, such that both
$\phi$ and  $\psi$ are in the domains of the operators $Q$ and $P$, are differentiable and vanish at
infinity. Then,
\beano
(\mathcal W X )(x,y) & = & \frac 1{(2\pi)^{\frac 12}}\;\text{Tr}\;[U(x,y)^* X] = \frac 1{(2\pi)^{\frac 12}}\;
     \langle U(x,y)\psi\mid \phi\rangle_\h \\
     & = & \frac 1{(2\pi)^{\frac 12}}\;\int_{\mathbb R}e^{ix (\xi - \frac y2 )}\;\overline{\psi (\xi -y )}\;\phi(\xi )\;
         d\xi\; .
\enano
Thus,
\beano
(\mathcal W Q\vee I_\h (X) )(x,y) & = & \frac 1{(2\pi)^{\frac 12}}\;\langle U(x,y)\psi\mid Q\phi\rangle_\h \\
   & = & \frac 1{(2\pi)^{\frac 12}}\;\int_{\mathbb R}e^{ix (\xi - \frac y2 )}\;
        \overline{\psi (\xi -y )}\;\xi \phi(\xi )\;d\xi \\
    & = & \left( -i\frac \partial{\partial x} + \frac y2\right)\left[\frac 1{(2\pi)^{\frac 12}}
     \;\int_{\mathbb R}e^{ix (\xi - \frac y2 )}\;\overline{\psi (\xi -y )}\;\phi(\xi )\;d\xi\right]\; ,
\enano
implying
$$ (\mathcal W Q\vee I_\h (X ))(x,y) = \left( -i\frac \partial{\partial x} + \frac y2\right)
(\mathcal W X )(x,y)\; .$$
Extending by linearity on appropriate domains, we get
$$
  \mathcal W Q\vee I_\h \mathcal W^{-1} = -i\frac \partial{\partial x} + \frac y2 = Q_1\; . $$

Next,
\beano
(\mathcal W P\vee I_\h (X ))(x,y) & = & \frac 1{(2\pi)^{\frac 12}}\;\langle U(x,y)\psi\mid P\phi\rangle_\h \\
   & = & \frac 1{(2\pi)^{\frac 12}}\;\int_{\mathbb R}e^{ix (\xi - \frac y2 )}\;
        \overline{\psi (\xi - y )}\;\left( -i\frac \partial{\partial\xi} \right) \phi(\xi )\;d\xi \; .
\enano
Now,
\beano
-i\frac \partial{\partial\xi}\; \left[ e^{ix(\xi - \frac y2)}\;\overline{\psi (\xi - y )}\; \phi (\xi )\right]
 & = & x\;e^{ix(\xi - \frac y2)}\;\overline{\psi (\xi - y )}\; \phi (\xi ) \\
 & \quad & + e^{ix(\xi - \frac y2)}\;\left(-i\frac \partial{\partial\xi}\right)\overline{\psi (\xi - y )}\; \phi (\xi )\\
  & \quad &  +  e^{ix(\xi - \frac y2)}\;\overline{\psi (\xi - y )}\;
   \left(-i\frac \partial{\partial\xi}\right)\phi (\xi )\; .
 \enano
 Integrating both sides of this equation with respect to $\xi$ from $-\infty$ to $\infty$ and noting that
$\psi (\xi ), \; \phi (\xi ) \longrightarrow 0$ as $\xi \longrightarrow \pm \infty$, and
$\displaystyle{\frac \partial{\partial\xi} \psi (\xi - y ) = -\frac \partial{\partial y} \psi (\xi - y )}$,
we get,
\beano
0
 & = & \frac x{(2\pi)^{\frac 12}}\;\int_{\mathbb R}  e^{ix(\xi - \frac y2)}\;\overline{\psi (\xi - y )}\; \phi (\xi )
  + \frac 1{(2\pi)^{\frac 12}}\;\int_{\mathbb R}e^{ix(\xi - \frac y2)}\;
   \left(i\frac \partial{\partial y}\right)\overline{\psi (\xi - y )}\; \phi (\xi )\\
  & \quad &  + \frac 1{(2\pi)^{\frac 12}}\;\int_{\mathbb R} e^{ix(\xi - \frac y2)}\;\overline{\psi (\xi - y )}\;
   \left(-i\frac \partial{\partial\xi}\right)\phi (\xi )\; .
 \enano
Thus,
$$ (\mathcal W P\vee I_\h (X ))(x,y) = \left( -i\frac \partial{\partial y} - \frac x2\right)
(\mathcal W X )(x,y)\; ,$$
and again, extending by linearity on appropriate domains we get
$$
  \mathcal W P\vee I_\h \mathcal W^{-1} = -i\frac \partial{\partial x} + \frac y2 = P_1\; . $$

\subsection{Proof of (\ref{polar-dec})-(\ref{antiunit-map3})}
Since the vectors $\Psi_{jk} , \; j,k = 0,1,2, \ldots , \infty$, form a basis of $\htil \;
( = L^2 (\mathbb R^2 , dxdy ))$, we may write
$$ U_1 (x,y)\Phi_\beta = \sum_{i=0}^\infty \lambda_i^{\frac 12} U_1 (x,y ) \Psi_{ii}
    = \sum_{i,j,k = 0}^\infty \lambda_i^{\frac 12}\langle \Psi_{jk}\mid U_1 (x,y )\Psi_{ii}\rangle_{\htil}\;
       \Psi_{jk}\; . $$
Now, using the isometry $\mathcal W \phi_{jk} = \mathcal W (\vert\phi_j\rangle\langle \psi_k \vert) =  \Psi_{jk}$
(see (\ref{hilb-sch-bas}) and (\ref{alg-transf2})), the first relation in (\ref{bi-alg1}) and the fact that the
vectors $\phi_i , \; i=0,1,2, \ldots , \infty$, form an orthonormal basis of $\h$, we obtain
\beano
\langle \Psi_{jk}\mid U_1 (x,y )\Psi_{ii}\rangle_{\htil} & = & \text{Tr} \left[ \vert\phi_k\rangle\langle\phi_j \vert
    U(x,y)\vert\phi_i\rangle\langle\phi_i\vert\right] = \langle\phi_j \mid U(x,y)\phi_i\rangle\; \delta_{ik}\\
    & = & (2\pi)^{\frac 12}\; \overline{\Psi_{ji} (x,y)}\; \delta_{ik}\; .
\enano
Thus,
\be
  U_1 (x,y)\Phi_\beta = (2\pi)^{\frac 12} \sum_{i,j = 0}^\infty \lambda_i^{\frac 12}
    \;\overline{\Psi_{ji}(x,y)}\;\Psi_{ji}\; .
\label{antilineq1}
\end{equation}
Similarly,
\be
  U_1 (x,y)^*\Phi_\beta = (2\pi)^{\frac 12} \sum_{i,j = 0}^\infty \lambda_i^{\frac 12}
    \;\Psi_{ij}(x,y)\;\Psi_{ji} = (2\pi)^{\frac 12} \sum_{i,j = 0}^\infty \lambda_j^{\frac 12}
    \;\Psi_{ji}(x,y)\;\Psi_{ij}\; .
\label{antilineq2}
\end{equation}
Next, applying the operator $S_\beta$ to both sides of (\ref{antilineq1}) and taking account of the
fact that this operator is antilinear, we get
$$ S_\beta \left[ U_1 (x,y)\Phi_\beta\right] = U_1 (x,y)^*\Phi_\beta =
    (2\pi)^{\frac 12} \sum_{i,j = 0}^\infty \lambda_i^{\frac 12}
    \;\Psi_{ji}(x,y)\;S_\beta\Psi_{ji}\; $$
Comparing this equation with (\ref{antilineq2}) we immediately see that
$$  S_\beta \Psi_{ji} = \left[\frac {\lambda_j}{\lambda_i}\right]^\frac 12 \Psi_{ij}\;, $$
from which (\ref{polar-dec})-(\ref{antiunit-map3}) follow directly.

\section*{Acknowledgements}
This work was partially supported by the Ministero Affari
Esteri, Italy, through its program of financial support for international
cooperations, Bando CORI 2003, cap. B.U. 9.3.0001.0001.0001, and
through grants from the Natural Sciences and Engineering Research
Council (NSERC), Canada and the Fonds qu\'eb\'ecois de la recherche
sur la nature et les technologies (FQRNT), Qu\'ebec.


\begin{thebibliography}{99}
\bibitem{agtar} G.S. Agarwal and K. Tara, {\em Nonclassical properties of states generated by the
     the exitations on a coherent state\/}, Phys. Rev. {\bf A43}, (1991), 492-497.

\bibitem{AEG} S.T. Ali, M. Engli\v s and J.-P. Gazeau, {\em Vector coherent states from
       Plancherel's theorem, Clifford algebras and matrix domains\/}, J. Phys. {\bf A37},
       (2004), 6067-6089.

\bibitem{AAG-book} S. T. Ali, J.-P. Antoine and J.-P. Gazeau, {\em Coherent States, Wavelets and
their Generalizations\/}, Springer-Verlag, New York (2000).

\bibitem{aliemch} S.T. Ali and G.G. Emch, {\em Geometric quantization: Modular reduction theory and
   coherent states\/}, J. Math. Phys. {\bf 27}, (1986), 2936-2943.

\bibitem{bagant} J.P. Antoine, F. Bagarello, {\em  Localization properties
and wavelet-like orthonormal bases for the lowest Landau level},
in {\em Advances in Gabor Analysis}, H.G. Feichtinger, T. Strohmer
Eds., Birkh\"auser, Boston, 2003


\bibitem{Aron} A. Aronszajn, {\em Theory of reproducing kernels\/}, Trans. Amer. Math. Soc.
{\bf 66}, (1950),  337-404.

\bibitem{BMS} F. Bagarello, G. Morchio and F. Strocchi, {\em Quantum Corrections to the
Wigner Crystal.  An Hartree-Fock Expansion}, Phys. Rev. B, {\bf
48}, (1993), 5306-5314.

\bibitem{fox} R.F. Fox, M. H. Choi, {\em Generalized coherent states for systems
with degenerate energy spectra}, Phys. Rev. A., {\bf 64}, (2001), 02104-1 -- 02104-6.

\bibitem{GK} J.P. Gazeau, J.R. Klauder, {\em Coherent states for systems with discrete and continuous
spectrum}, J. Phys. A, {\bf 32}, (1999), 123-132.

\bibitem{gaznov} J.P. Gazeau and M. Novaes, {\em Multidimensional generalized coherent states\/},
      J. Phys. A, {\bf 36}, (2003), 199-212.

\bibitem{hahuwi} R. Haag, N. Hugenholtz and M. Winnink, {\em On the equilibrium states in quantum statistical
   mechanics\/}, Commun. Math. Physics {\bf 5},  (1967), 215-236.

\bibitem{klau} J. R. Klauder, {\em Coherent states for the hydrogen atom\/}, J. Phys. {\bf A29}, (1996),
  L293-L298.

\bibitem{Mesch} H. Meschkowsky, {\em Hilbertsche R\"aume mit Kernfunktionen\/}, Springer-Verlag, Berlin
(1962).

\bibitem{simon} D.T. Simon, M.R. Geller, {\em Electron-phonon dynamics in an ensemble of
 nearly isolated nanoparticles}, Phys. Rev. B, {\bf 64}, (2001),
 224504

 \bibitem{tak1} M. Takesaki, {\em Theory of Operator Algebras. I\/}, Springer-Verlag, New York (1979).

\bibitem{tak2} M. Takesaki, {\em Tomita's Theory of Modular Hilbert Algebras and its Applications\/}, Lect.
   Notes in Math., Vol. 128, Springer-Verlag, Berlin (1970).

 \bibitem{sanali} K. Thirulogasanthar and S.T. Ali, {\em A class of vector coherent states defined
 over matrix domains\/}, J. Math. Phys. {\bf 44},  (2003), 5070-5083.



\end{thebibliography}
\end{document}